\documentclass[preprint]{aastex631}

\usepackage[version=4]{mhchem}
\usepackage{array}
\usepackage{lipsum}
\usepackage{soul}
\usepackage{wrapfig}
\usepackage{multirow}
\usepackage{amsmath}
\usepackage{hyperref}
\usepackage{textcomp}
\usepackage{gensymb}
\usepackage{longtable, booktabs, threeparttablex}

\makeatletter
\DeclareRobustCommand{\HII}{%
  \mbox{H\check@mathfonts\fontsize\sf@size\z@\selectfont II}%
}
\makeatother

\begin{document}

\title{Spatial and Chemical Complexity in the W75N Star-Forming Region}

\author[0009-0007-3950-335X]{Morgan M. Giese}
\affiliation{Department of Astronomy, University of Wisconsin-Madison \\
475 N Charter St, Madison, Wisconsin 53706, USA}

\author[0000-0002-1353-2562]{Will E. Thompson}
\affiliation{Department of Chemistry, University of Wisconsin-Madison \\
1101 University Ave, Madison, Wisconsin 53706, USA}

\author[0000-0002-0500-4700]{Dariusz C. Lis}
\affiliation{Jet Propulsion Laboratory, California Institute of Technology \\
4800 Oak Grove Drive, Pasadena, CA, 91109, USA}

\author[0000-0001-6015-3429]{Susanna L. Widicus Weaver}
\affiliation{Department of Astronomy, University of Wisconsin-Madison \\
475 N Charter St, Madison, Wisconsin 53706, USA}
\affiliation{Department of Chemistry, University of Wisconsin-Madison \\
1101 University Ave, Madison, Wisconsin 53706, USA}

\begin{abstract}

We present the analysis of NOEMA interferometric observations of the high-mass star-forming region W75N(B) with a focus on molecular composition and distribution of prebiotic molecules in the source's multiple cores. Over twenty molecules are identified across the region, with many being fit for column density, rotational temperature, spectral line full width half maximum, and v$_{lsr}$. This work includes the first known detection and initial analysis of complex organic molecules in the MM2 and MM3 regions. Furthermore, parameter maps were created from the six molecules that were well fit across multiple regions. The molecular emission was imaged and correlated across different molecules and the continuum to reveal structural features. From the spatial and spectral analysis of the MM1 region, these results concur with those from other studies showing that there is a difference in chemical composition between the MM1a and MM1b regions, with sulfur-bearing molecules tracing MM1a and organic molecules tracing MM1b. The molecular emission imaged toward the MM3 region reveals two peaks, possibly indicating the presence of multiple young stellar objects. These results provide detailed quantitative information about the physical parameters and distributions of molecules in this source. Additionally, these results are part of a follow-up of a single-dish survey of multiple star-forming regions and are discussed in this context.

\end{abstract}


\section{Introduction} \label{sec:intro}

Broadband line surveys provide a means to detect and quantify the molecular compositions of interstellar sources.  For star-forming regions in particular, such surveys enable  identification and analysis of complex organic molecules (COMs), defined as having more than six atoms, that could serve as precursors for life. COMs can be found in abundance in star-forming regions and protoplanetary disks, and these molecules play a role in the formation of planetary systems \citep{Blake1998, VanDishoeck2006}. Many of these prebiotic molecules are generally formed from the photo and thermal processing of icy dust grains, from which they ultimately sublimate into the gas phase \citep{MunozCaro2002, Oberg2011, Altwegg2019}. Once in the gas phase, interferometric studies become essential for probing the chemical composition and distribution of molecules in different star-forming regions. Interferometers such as the NOrthern Extended Millimeter Array (NOEMA) and the Atacama Large Millimeter/Submillimeter Array (ALMA) allow for targeted observations with high spatial and spectral resolution to characterize these chemically-rich environments \citep{Jorgensen2004, Tobin2011, Tycho2021}.

W75N(B) is a high-mass star-forming region in the Cygnus X molecular cloud located at a distance of $\sim$2 kpc with a v$_{lsr}$ of 10 km s$^{-1}$ \citep{haschick_vlbi_1981, Minh2010, van_der_walt_protostellar_2021}. W75N(B) (hereafter simply  W75N) is associated with three millimeter continuum cores MM1, MM2, and MM3; the brightest core MM1 is further categorized into the sub-regions MM1a and MM1b \citep{hunter_water_1994, shepherd_clustered_2003, Minh2010, van_der_walt_protostellar_2021}. In addition to these sub-regions, three ultra-compact {\HII} regions VLA1, VLA2, and VLA3 are detected, with VLA3 being cospatial with MM1a \citep{hunter_water_1994, torrelles_evidence_2003, shepherd_clustered_2003}. The {\HII} regions are associated with multiple masers: VLA1 and VLA2 with OH, \ce{H2O}, and \ce{CH3OH} masers and VLA3 with a \ce{H2O} maser \citep{torrelles_evidence_2003, minier_vlbi_2001, shepherd_clustered_2003, hutawarakorn_oh_2002, fish_fullpolarization_2005, surcis_methanol_2009}.

Recent chemical studies of W75N include observations with the Submillimeter Array (SMA) in the 316 to 329 GHz window \citep{van_der_walt_protostellar_2021};  with the Caltech Submillimeter Observatory (CSO) in the 220 to 267 GHz window \citep{WidicusWeaver2017}; and  with the SMA in the 215 and 345 GHz windows \citep{Minh2010}.  These observations all focused on the chemical composition of the MM1 region. It was determined that MM1a and MM1b display distinct spatial morphologies of organic and sulfur-bearing molecules. However, none of these studies provide detailed analysis on the molecular composition of the MM2 or MM3 regions. To add further information to these previous studies, presented here is an analysis of the molecular composition of the MM1a, MM1b, MM2, and MM3 regions along with a characterization of the spatial distribution of the discovered molecules. Below we present the details of our NOEMA observations of W75N, the spectral and spatial results of these observations for each molecule detected, and a discussion of the implications of these results in the context of COM formation in star-forming cores. As this work is part of a follow-up NOEMA survey of multiple star-forming cores originally analyzed in \cite{WidicusWeaver2017} using the CSO, molecules identified herein are also directly compared with those identified in sources previously imaged with NOEMA.

\section{Observations}

W75N was observed using the IRAM/NOEMA interferometer\footnote{IRAM is supported by INSU/CNRS (France), MPG (Germany), and IGN (Spain).} with the A configuration for $\sim$8 hours on February 13 and 28 2022. The pointing center of the observations was $\alpha$(J2000) = 20$^{h}$38$^{m}$36.25$^{s}$, $\delta$(J2000) = 42$^{o}$37\arcmin33\arcsec.0. The observations were centered at a local oscillator frequency of 145.250 GHz with frequency coverage from 127.823 - 135.311 GHz and 143.116 - 150.666 GHz for the lower and upper sidebands, respectively. 

The baselines of the array ranged from 32.0 m to 920.0 m in the 12A configuration. The data were obtained using the PolyFiX correlator, which gave a channel spacing of 2.0 MHz across both sidebands. In addition, 28 high-resolution spectral windows were selected within the two sidebands with channel spacings of 62.5 kHz (0.13 km s$^{-1}$). Bandpass calibration was performed using observations of 2013+370, and flux calibration was performed using observations of MWC349 and 2010+723. The maximum recoverable scale was 8\arcsec.0, which was calculated using Equation 3.28 in \cite{Cortes}. Phase and amplitude calibration was performed using 2013+370 and J2050+363. The system temperatures varied between 60 K and 180 K during the observations. A summary of the observational setup is shown in Appendix \ref{sec:windowspecs}. 

\begin{figure*}
    \centering
    \includegraphics[width = \textwidth]{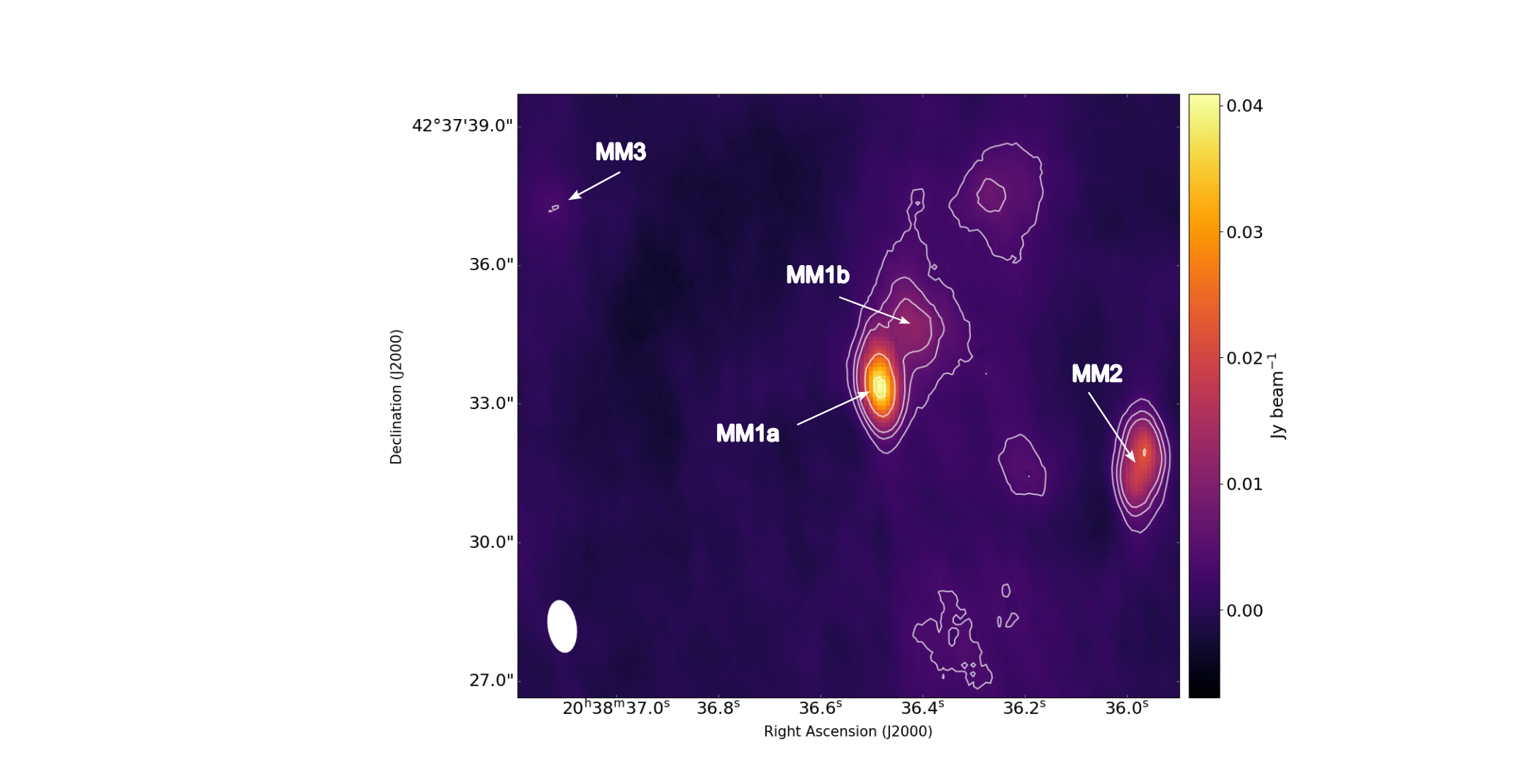}
    \caption{130 GHz continuum emission of the MM1, MM2, and MM3 regions. White contour levels are at 1, 2, 3, 7, and 12 times $\sigma$ ($\sigma$ = 3.079 mJy beam$^{-1}$). The synthesized beam size is shown in the lower left corner.}
    \label{fig:continuum}
    
\end{figure*}

Each dataset was reduced and cleaned following a similar procedure to that described in \cite{Thompson2023} using the GILDAS packages CLIC and MAPPING\footnote{http://www.iram.fr/IRAMFR/GILDAS}. This resulted in a synthesized beam of $\sim$ 1.00\arcsec ~$\times$ 0.53\arcsec ~(PA = --171.79\degree) in the lower sideband and $\sim$ 0.82\arcsec ~$\times$ 0.42\arcsec ~(PA = 8.8\degree) in the upper sideband. To clean each of the 28 high-resolution datacubes, a robust weighting system was used until the maximum amplitude of the absolute value of the residual equaled twice the noise. Each datacube was then continuum-subtracted using the STATCONT software package \citep{SanchezMonge2018}. STATCONT determines the continuum level at each individual pixel utilizing a user-inputted RMS value for each datacube. Appendix \ref{sec:windowspecs} reports the RMS values of each cube. Figure \ref{fig:continuum} shows the imaged continuum at 130 GHz. The MM1a, MM1b, MM2, and MM3 cores are labeled in accordance with previous works \citep{Minh2010, van_der_walt_protostellar_2021}. Compared to these studies, the continuum emission reported here is at a much lower signal-to-noise ratio. This is especially evident in the MM3 region, where only a 1-$\sigma$ contour is seen. The flux ratio between the MM3 and MM1 regions is roughly 1/8. This is comparable to flux ratios seen in other works \citep{Minh2010, zeng_submillimeter_2023}. Additionally, an extracted spectrum from the MM3 position listed in \cite{zeng_submillimeter_2023} yielded no detectable lines above 3$\sigma$. Also unlike these previous studies, molecular emission is detected from all four regions and is analyzed herein.

\section{Line Identification and Analysis}

The Global Optimization and Broadband Analysis Software (GOBASIC) was used to analyze each molecule in the broadband spectra. GOBASIC performs a broadband spectral analysis for multiple molecules simultaneously under the assumptions of Gaussian line shapes and local thermodynamic equilibrium (LTE) for a given molecule. For each molecule, GOBASIC derives column density in cm$^{-2}$, rotational temperature in K, spectral line full width half maximum ($\Delta v$) in km s$^{-1}$, and v$_{lsr}$ in km s$^{-1}$. The intricacies of  GOBASIC operation are provided in \cite{Rad2016}, but a brief description is provided here.

\begin{figure*}[t]
  \includegraphics[width=0.9\textwidth]{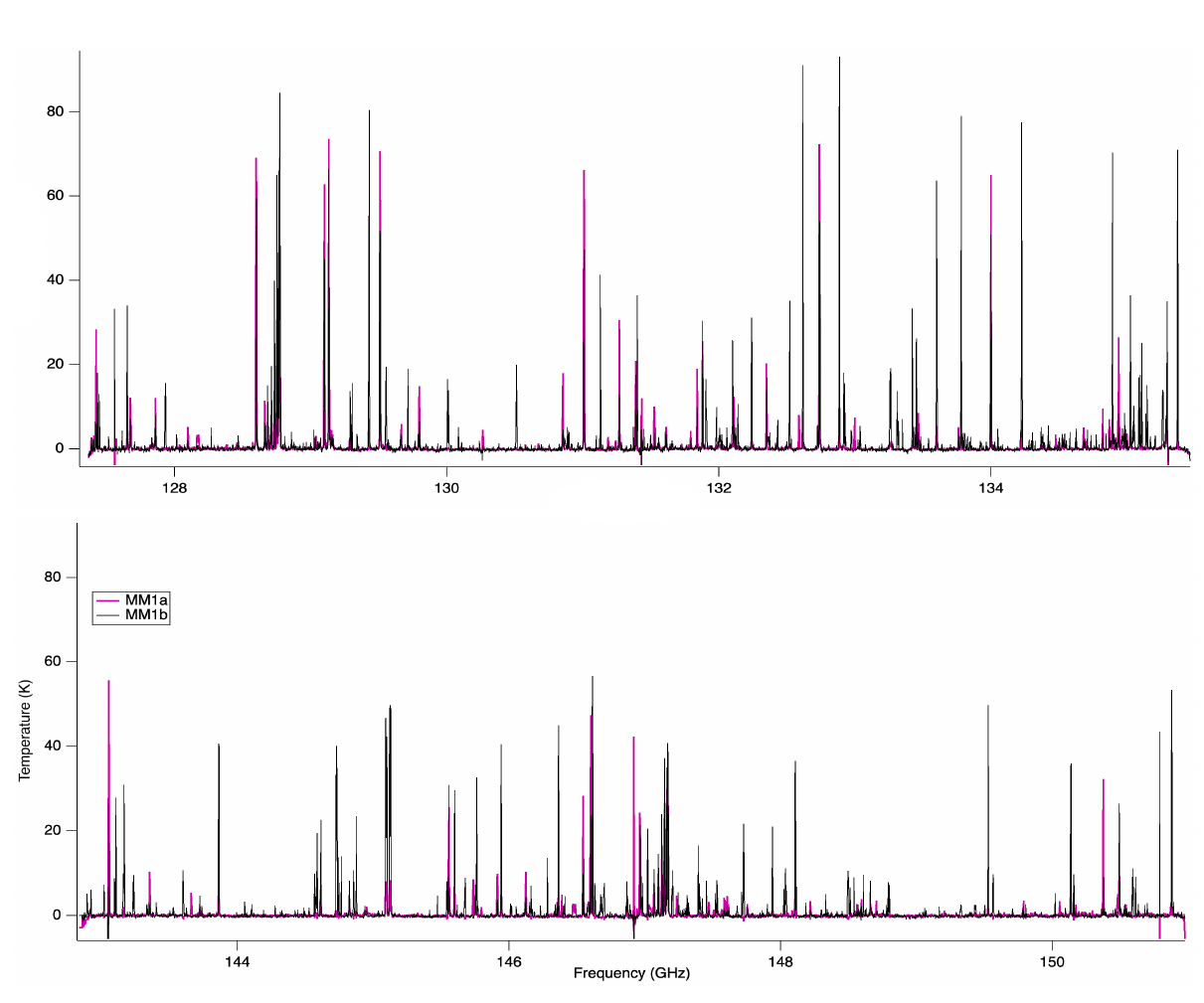}
  \caption{Spectra for MM1a and MM1b for the lower sideband (top) and the upper sideband (bottom).
  \label{fig:MM1_spec}}
\end{figure*}

\begin{figure*}[t]
  \includegraphics[width=0.9\textwidth]{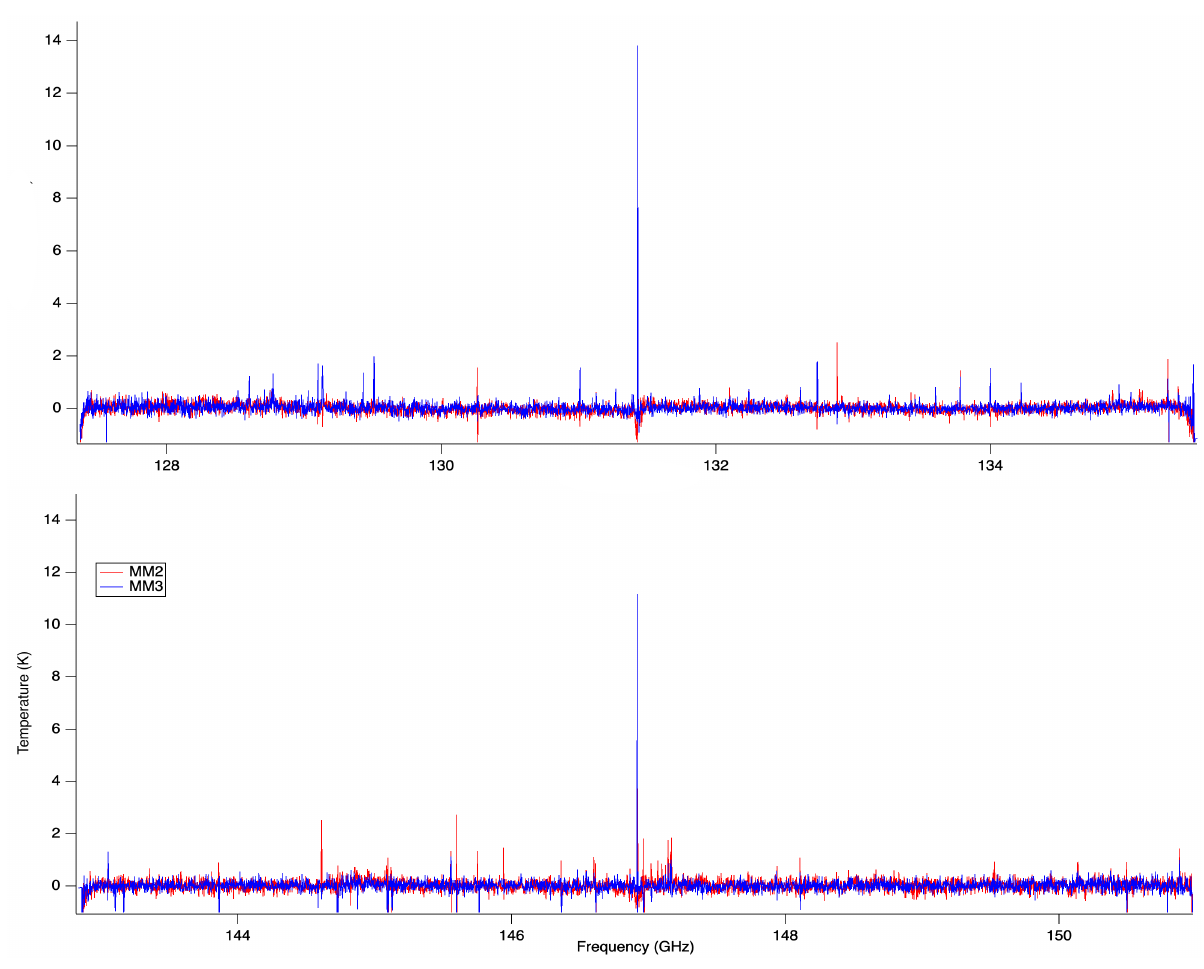}
  \caption{Spectra for MM2 and MM3 for the lower sideband (top) and the upper sideband (bottom).
  \label{fig:MM23_spec}}
\end{figure*}

For the analysis presented herein, GOBASIC is used in a similar manner to that in previous works \citep{WidicusWeaver2017, Zou2017, Wright2022, Thompson2023, giese_mapping_2023}. GOBASIC utilizes molecular catalog files from both the Cologne Database for Molecular Spectroscopy (CDMS) \citep{Muller2001, Muller2005} and the Jet Propulsion Laboratory (JPL) Spectral Line Catalog \citep{Picket1998} to perform a full broadband analysis of multiple molecules by comparing the provided catalog information to the observational spectra. Specifics on which database catalog was used for each molecule can be found in \cite{WidicusWeaver2017}. From a user-inputted initial guess for each physical parameter and the partition function information from the catalog, GOBASIC creates a spectral simulation to compare to the observational data. The partition function used in the program follows the general form 
\begin{equation}
Q(T) = \alpha T^{\beta}[\gamma + exp (\epsilon/T)].
\end{equation}
For most molecules identified in W75N, the coefficients for the partition function interpolation and the process by which they were calculated were previously reported in \cite{WidicusWeaver2017}. Discussion on the exception to this can be found below. The values for column density, temperature, line width, and velocity were also constrained to physically-meaningful values. Furthermore, the emission is assumed to be resolved and fill the small synthesized beam. Therefore, no correction for source size was made. At the beginning of the analysis process, target molecules were chosen based on previously-published line surveys of W75N \citep{Minh2010, WidicusWeaver2017, van_der_walt_protostellar_2021}. Any spectral lines not accounted for by molecules from these surveys were then cross-referenced with other molecular catalogs, resulting in new molecules being added to the analysis. To analyze the full region using this method, a spectrum was extracted from the pixel with peak continuum flux in each core, described in Table \ref{tab:pixel}. The spectra for MM1a and MM1b can be seen in Figure \ref{fig:MM1_spec}. Spectra for MM2 and MM3 can be seen in Figure \ref{fig:MM23_spec}. Hundreds of molecular transitions were present across all four regions. As all of these transitions from all molecules were fit simultaneously using GOBASIC, a detailed analysis of each individual transition, commonly seen in other studies of star-forming regions, was not performed. All possible transitions listed in the available catalog for each molecule were included in this analysis, probing an upper energy range of 7 K to 7200 K. While not all of these transitions were visible in the low-resolution datacubes, the most intense transitions were seen in both the low-resolution and the 28 high-resolution datacubes. For these high-intensity transitions, the upper energies are provided in Table \ref{tab:hires_lines}.

\begin{table*}
  \centering
  
    \caption{\label{tab:pixel}Pixel locations with peak continuum flux used for spectral extraction.}

    \begin{tabular}{c|c|c} 
        \hline
        Continuum Core & Right Ascension (J2000) & Declination (J2000)\\
        \hline
        MM1a    &  20$^{h}$38$^{m}$36.5$^{s}$   & 42$^{o}$37\arcmin33\arcsec.3\\
        MM1b    &  20$^{h}$38$^{m}$36.4$^{s}$   & 42$^{o}$37\arcmin34\arcsec.6\\
        MM2    &  20$^{h}$38$^{m}$36.0$^{s}$   & 42$^{o}$37\arcmin31\arcsec.9\\
        MM3    &  20$^{h}$38$^{m}$37.1$^{s}$   & 42$^{o}$37\arcmin37\arcsec.2\\

        \hline
    \end{tabular}

\end{table*}

More than twenty molecules were detected in MM1a with at least three lines above 3$\sigma$: \ce{CH3CN} (ethyl cyanide), \ce{CH3OH} (methanol), \ce{CS} (carbon monosulfide), C$^{33}$S, C$^{34}$S, \ce{DCN} (deuterium cyanide), \ce{H2CO} (formaldehyde), \ce{H2CS} (thioformaldehyde), \ce{HC3N} $v_7$ (cyanoacetylene), H$^{13}$CCCN, HC$^{13}$CCN, HCC$^{13}$CN, \ce{t-HCOOH} (formic acid), \ce{HDO} (deuterated water), \ce{HNCO} (isocyanic acid), \ce{NH2CHO} (formamide), \ce{NO} (nitric oxide), \ce{OCS} (carbonyl sulfide), \ce{SiO} (silicon monoxide), \ce{SO} (sulfur monoxide), $^{33}$SO, S$^{18}$O, \ce{SO2} (sulfur dioxide), $^{33}$SO$_2$, and $^{34}$SO$_2$.

As DCN was not originally reported by  \cite{WidicusWeaver2017}, it was necessary to create a new interpolation of the partition function for its coefficients. The process of doing so is described in \cite{WidicusWeaver2017}. The coefficients for interpolation, along with the line catalog database and version, are summarized in Table \ref{tab:DCN}. Similar to other molecules reported in \cite{WidicusWeaver2017}, the relative deviation (defined in Equation \ref{eqn:2}) between the new interpolated partition function values and those reported in the catalog is also provided in the table.

\begin{equation}
\label{eqn:2}
   \delta = Q_{interpolated}/Q_{catalog} -1 
\end{equation}

Physical parameters were obtained for \ce{CH3CN}, \ce{CH3OH}, \ce{HC3N} $v_7$, \ce{HNCO}, \ce{NH2CHO}, \ce{OCS}, \ce{SO}, \ce{SO2}, and $^{34}$SO$_2$.  These molecules all have numerous lines in the spectrum, enabling a full analysis. The calculated physical parameters for these molecules are shown in Table \ref{tab:mm1a-table}. For other molecules detected in this region, it was difficult to derive accurate parameters via fitting due to each molecule having less than five observable lines over 3-$\sigma$ that were not affected by blending.

These values for molecules in MM1a are similar to those determined by \cite{van_der_walt_protostellar_2021}. Generally, the column densities are all of the same order of magnitude except for \ce{CH3OH}, which is lower in the results shown here. However, when comparing the column density of \ce{CH3OH} to that reported in \cite{Minh2010}, our results match. {\bf Some methanol lines are under-predicted by the best fit, indicating that features from other molecules were blended with these lines or additional temperature components were present for methanol. 
Neither of these cases can be fully quantified without additional information and therefore were not included in the analysis presented here. }
 Apart from \ce{HC3N} and the sulfur-bearing molecules, the temperatures reported here are higher than those in \cite{van_der_walt_protostellar_2021}. The FWHM values for most molecules are also generally higher than those previously reported, although the FWHM of \ce{SO2} is comparable. Velocity values are also comparable. The results presented here are from an analyzed spectrum located closer to the center of the MM1a core than that of \cite{van_der_walt_protostellar_2021},  and this could be the cause of any discrepancies.

\begin{table*}
\centering
  
    \caption{\label{tab:DCN}Partition function coefficient interpolation for DCN. Deviation between the interpolated partition function values and the catalog tabulated partition function values is listed as a range of $[\delta_{min},\delta_{max}]$ in units of percent.}

    \begin{tabular}{|c|c|c|c|c|c|c|} 

        \hline
        $\alpha$ & $\beta$ & $\gamma$ & $\epsilon$ & $\delta(\%)$\tnote{d} & Database & Version \\
        \hline
        1.727 & 1 & 0 & 0.5836 & [0.00030 0.048] & JPL & 2006 Apr\\

        \hline

    \end{tabular}

    \end{table*}
    
\begin{table*}
  \centering
  
    \caption{\label{tab:mm1a-table}Calculated physical parameters for MM1a from GOBASIC fits. The four parameters are column density in cm$^{-2}$ (N$_{\text{T}}$), temperature in K, spectral line full width half maximum in km s$^{-1}$ (FWHM), and v$_{lsr}$ in km s$^{-1}$ (v$_{lsr}$).}

    \begin{tabular}{|c|c|c|c|c|} 

        \hline
        \multicolumn{5}{|c|}{MM1a Derived Parameters} \\
        \hline
        Molecule & N$_{\text{T}}$ (cm$^{-2}$) & Temperature (K) & FWHM (km s$^{-1}$) & v$_{lsr}$ (km s$^{-1}$) \\
        \hline
        \ce{CH3CN} & 1.48(0.11)e+16 & 223.8(14.5) & 9.44(0.19) & 9.97(0.09)\\
        \ce{CH3OH} & 8.38(0.82)e+16 & 158.0(10.4) & 9.32(0.52) & 10.84(0.27)\\
        \ce{HC3N} $v_7$ & 2.26(0.83)e+15 & 27.9(8.8) & 10.23(0.67) & 9.69(0.30)\\
        \ce{HNCO} & 3.47(0.26)e+16 & 196.3(12.7) & 8.61(0.27) & 10.89(0.14)\\
        \ce{NH2CHO} & 1.32(0.57)e+15 & 91.5(39.3) & 8.88(1.03) & 9.37(0.53)\\
        \ce{OCS} & 2.42(0.97)e+16 & 13.9(2.2) & 8.82(1.17) & 10.65(0.56)\\
        \ce{SO} & 3.92(1.30)e+17 & 67.7(1.0) & 6.00(0.30) & 9.70(0.05)\\
        \ce{SO2} & 3.07(0.05)e+17 & 64.2(0.4) & 6.70(0.04) & 10.07(0.02)\\
        $^{34}$SO$_2$ & 2.77(0.45)e+16 & 144.2(22.4) & 8.60(0.46) & 9.80(0.21)\\

        \hline

    \end{tabular}

\end{table*}

More than twenty molecules were also detected in MM1b with at least three lines above 3$\sigma$: \ce{CH3CH2CN} (ethyl cyanide), \ce{CH3CN}, \ce{CH3OCH3} (dimethyl ether), \ce{CH3OH}, \ce{CS}, C$^{33}$S, C$^{34}$S, \ce{DCN}, \ce{H2CO}, \ce{H2CS}, \ce{HC3N} $v_7$, H$^{13}$CCCN, HC$^{13}$CCN, HCC$^{13}$CN, \ce{HCOOCH3} (methyl formate), \ce{HDO}, \ce{HNCO}, \ce{NH2CHO}, \ce{NO}, \ce{OCS}, \ce{SiO}, \ce{SO}, S$^{18}$O, \ce{SO2}, and $^{34}$SO$_2$. Physical parameters were obtained for \ce{CH3CN}, \ce{CH3OCH3}, \ce{CH3OH}, \ce{HCOOCH3}, \ce{HNCO}, and \ce{SO2}.  These molecules each have numerous lines in the spectrum, enabling a full analysis. The calculated physical parameters for these molecules are shown in Table \ref{tab:mm1b-table}. As in the MM1a region, it was difficult to derive accurate parameters via fitting for other molecules detected in this region due to each molecule having less than five observable lines over 3-$\sigma$ that were not affected by blending.

Similar to the MM1a region, all column density values of molecules in MM1b are consistent with those reported in \cite{van_der_walt_protostellar_2021}. In terms of temperature, \ce{CH3OCH3} and HNCO are also similar to the previously reported values. \ce{CH3OH} is considerably warmer with a temperature of 205 K compared to the previously reported value of 140 K, while \ce{CH3CN} and \ce{SO2} are colder. This is reasonable for all three molecules, as \ce{CH3OH} peaks in MM1b while \ce{CH3CN} and \ce{SO2} peak in MM1a \citep{Minh2010,van_der_walt_protostellar_2021}. Further discussion of molecular peaks can be found in Section \ref{sec:image}. As in MM1a, both FWHM and velocity values are consistent with those found by \cite{van_der_walt_protostellar_2021}.

\begin{table*}
  \centering
  
    \caption{\label{tab:mm1b-table}Calculated physical parameters for MM1b from GOBASIC fits. The four parameters are column density in cm$^{-2}$ (N$_{\text{T}}$), temperature in K, spectral line full width half maximum in km s$^{-1}$ (FWHM), and v$_{lsr}$ in km s$^{-1}$ (v$_{lsr}$).}

    \begin{tabular}{|c|c|c|c|c|} 

    \hline
    \multicolumn{5}{|c|}{MM1b Derived Parameters} \\
    \hline
    Molecule & N$_{\text{T}}$ (cm$^{-2}$) & Temperature (K) & FWHM (km s$^{-1}$) & v$_{lsr}$ (km s$^{-1}$) \\
    \hline
    \ce{CH3CN} & 3.26(0.22)e+16 & 56.7(0.9) & 4.32(0.07) & 8.52(0.04)\\
    \ce{CH3OCH3} & 3.82(0.18)e+17 & 110.3(5.3) & 5.95(0.22) & 8.60(0.14)\\
    \ce{CH3OH} & 1.17(0.02)e+18 & 204.7(2.7) & 6.80(0.08) & 8.25(0.04)\\
    \ce{HCOOCH3} & 7.28(0.52)e+16 & 89.4(5.4) & 5.79(0.27) & 8.59(0.12)\\
    \ce{HNCO} & 1.51(0.22)e+16 & 107.1(16.0) & 6.55(0.53) & 8.92(0.23)\\
    \ce{SO2} & 9.11(0.22)e+16 & 82.8(5.6) & 6.93(0.16) & 9.56(0.08)\\
    \hline

        \hline

    \end{tabular}

\end{table*}

Nine molecules were detected in MM2 with lines above 3$\sigma$: \ce{CH3CN}, \ce{CH3OCH3}, \ce{CH3OH}, CS, C$^{34}$S, \ce{HCOOCH3}, \ce{H2CS}, OCS, and SiO. However, these molecules have fewer than ten transitions visible above 3-$\sigma$ in the spectrum at the continuum peak for this core. Due to this, it was difficult to obtain an accurate fit for the physical parameters of any of them. To further analyze these molecules, fits were attempted in which temperatures were fixed to the values derived for MM1a and MM1b in order to obtain an estimation for column density. From this, only \ce{CH3OCH3} had a derived value for column density that was greater than the uncertainty. Fixed at 110 K, \ce{CH3OCH3} had a derived column density of 9.4(4.2)$\times10^{15}$ cm$^{-2}$. As the other molecules detected in MM2 had uncertainty values greater than the derived parameter values, even when fixing the temperature, they are not listed in a table here.

Seven molecules were detected in MM3 with lines above 3$\sigma$: \ce{CH3CN}, \ce{H2CS}, \ce{HCOOCH3}, \ce{HNCO}, \ce{OCS}, \ce{SO}, and \ce{SO2}. Physical parameters were obtained for \ce{CH3CN}, \ce{HNCO}, and \ce{SO2}.  These molecules each have numerous lines in the spectrum, enabling a full analysis. The calculated physical parameters for these molecules are shown in Table \ref{tab:mm3-table}. As in the MM1a and MM1b regions, it was difficult to derive accurate parameters via fitting for other molecules detected in this region due to each molecule having less than five observable lines over 3-$\sigma$ that were not affected by blending.

\begin{table*}
  \centering
  
    \caption{\label{tab:mm3-table}Calculated physical parameters for MM3 from GOBASIC fits. The four parameters are column density in cm$^{-2}$ (N$_{\text{T}}$), temperature in K, spectral line full width half maximum in km s$^{-1}$ (FWHM), and v$_{lsr}$ in km s$^{-1}$ (v$_{lsr}$).}

    \begin{tabular}{|c|c|c|c|c|} 

    \hline
    \multicolumn{5}{|c|}{MM3 Derived Parameters} \\
    \hline
    Molecule & N$_{\text{T}}$ (cm$^{-2}$) & Temperature (K) & FWHM (km s$^{-1}$) & v$_{lsr}$ (km s$^{-1}$) \\
    \hline
    \ce{CH3CN} & 8.98(5.09)e+13 & 112.4(53.3) & 3.53(0.85) & 13.28(0.54)\\
    \ce{HNCO} & 1.03(0.73)e+15 & 230.6(125.1) & 8.50(2.73) & 9.14(1.49)\\
    \ce{SO2} & 3.36(0.65)e+15 & 140.6(27.6) & 8.73(0.91) & 9.80(0.38)\\

    \hline

    \end{tabular}

\end{table*}

\section{Physical Parameter Maps}

In order to better understand how the chemistry changes across a star-forming region, maps of column density, temperature, and velocity were created for molecules that fit well in at least two regions. To do so, a spectrum for both the upper and lower sidebands was extracted from each pixel surrounding MM1a, MM1b, MM2, and MM3. For each specific pixel, the spectra from both sidebands were baseline-corrected with a first-order polynomial using the imodpoly algorithm of the pybaselines\footnote{https://github.com/derb12/pybaselines} python library and combined to create the finalized spectra to be analyzed using GOBASIC. GOBASIC was then run iteratively through all spectra to fit all molecules simultaneously. This procedure follows that used in \cite{giese_mapping_2023} and \cite{Thompson2023} with a key difference: pixels resulting in high uncertainties were not flagged and set to zero. Instead, any fits that hit the upper and lower bounds of the initial user-inputted parameter guesses were set to zero, while fits with high uncertainty were still included.

Maps were created for the six molecules with good fits in the MM1b region, as these were the only molecules that also fit well across MM2 and MM3. Specifically, maps were only created for a given molecule if that molecule had well-fit parameters described in the tables above. Overall, the MM2 and MM3 regions proved difficult to fit due to having a much lower signal-to-noise ratio than MM1, but there is still clear evidence of molecular emission in each region. These maps also show the first known evidence of complex organics in the MM2 region \citep{Minh2010,van_der_walt_protostellar_2021}. The parameter maps for \ce{CH3CN} are shown in Figure \ref{fig:CH3CN}, while the maps for all other molecules are shown in Appendix \ref{sec:maps}. Full uncertainty maps of each parameter across every pixel are shown in Appendix \ref{sec:appendix}. It is also important to note that in both the MM2 and MM3 regions, most pixels have uncertainties greater than the derived parameter values. The analysis of parameters across these two regions using this mapping method should therefore be seen as an initial estimate rather than a quantitative result. For many molecules, there is a clear difference in calculated values between the MM1a and MM1b regions. Furthermore, the parameter maps do not have a 1:1 correlation with the continuum contours. This is further discussed in Section \ref{sec:image}.

\begin{figure*}[t]
  \includegraphics[width=0.9\textwidth]{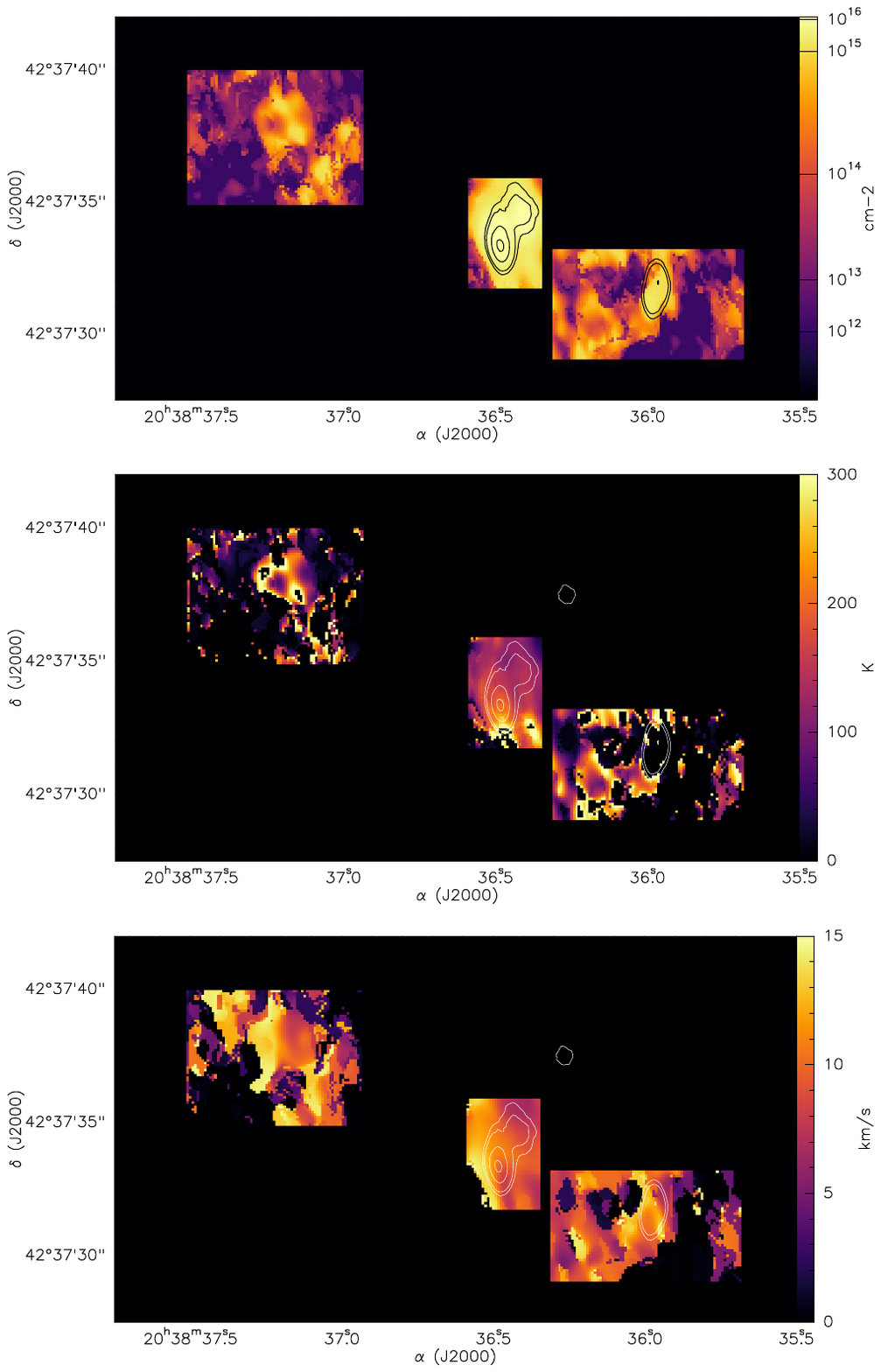}
  \caption{Parameter maps for \ce{CH3CN} across MM1, MM2, and MM3 (from top to bottom: column density, kinetic temperature, and v$_{lsr}$). Black and white contours correspond to the continuum levels at 2, 3, 7, and 12 times $\sigma$ ($\sigma$ = 3.079 mJy beam$^{-1}$). The colors of the contours are arbitrary and differ between maps purely for aesthetic purposes. In both the MM2 and MM3 regions, most pixels have uncertainties greater than the derived parameter values. The analysis of parameters across these two regions using this mapping method should therefore be seen as an initial estimate rather than a quantitative result.
  \label{fig:CH3CN}}
\end{figure*}

Generally, the column density for each molecule peaks at either MM1a or MM1b. \ce{CH3OCH3}, \ce{CH3OH}, and \ce{HCOOCH3} peak in MM1b, while \ce{SO2} peaks in MM1a. However for \ce{CH3CN} and HNCO, the column density is relatively uniform across both regions. \ce{CH3OH} has the highest column density out of all the molecules of 1.0$\times$10$^{18}$ cm$^{-2}$ in MM1b. In MM2 and MM3, all molecules are found to have column densities of approximately an order of magnitude less than those found for MM1. For \ce{CH3CN} and \ce{CH3OCH3}, the column density across MM2 is fairly uniform, whereas in \ce{CH3OH} and \ce{HCOOCH3} it increases southward across the continuum core. As is evident in the maps for \ce{CH3CN} and HNCO, MM3 is seemingly split into two parts, with the west side of the region having a higher column density than the east side. Conversely, the map for \ce{SO2} is much more uniform across both sides.

Overall, temperature proved much more difficult to fit in each region. Maps for MM2 and MM3 were most affected due to a considerable intensity difference between these regions and MM1. \ce{CH3CN} and \ce{CH3OH}  have the highest temperature values of around 280 K in MM1a and MM2, respectively. Unlike column density, most molecules have higher temperatures in MM1a than in MM1b. Out of the six molecules tested, \ce{CH3OH} is the only molecule with a higher temperature in MM1b. The temperature map of \ce{SO2} shows two areas of the MM1 region where values drop drastically. This may be due to \ce{SO2} having some optically thick lines as reported in \cite{van_der_walt_protostellar_2021}. While there were no optically thick lines found in this analysis, further results would be beneficial in determining the cause of these temperature drops.

The velocity maps show a distinct difference in values when comparing MM1a to MM1b. Across the entirety of MM1, every molecule shows a gradual increase in velocity from around 5.0 km s$^{-1}$ to 15.0 km s$^{-1}$ west to east. For the case of \ce{HCOOCH3}, velocity was poorly-derived in MM1a due to a much lower signal-to-noise. In MM2, all molecules show a velocity of around 10 km s$^{-1}$ inside the continuum contours. In MM3, the maps for \ce{CH3CN} and HNCO once again show the split of the region into east and west portions. Both molecules gradually increase in value from around 10.0 km s$^{-1}$ to 15.0 km s$^{-1}$ west to east. \ce{SO2}, however, is more uniform in MM3 with a value around 12.0 km s$^{-1}$.

\begin{figure*}[ht]
  \includegraphics[width=\textwidth]{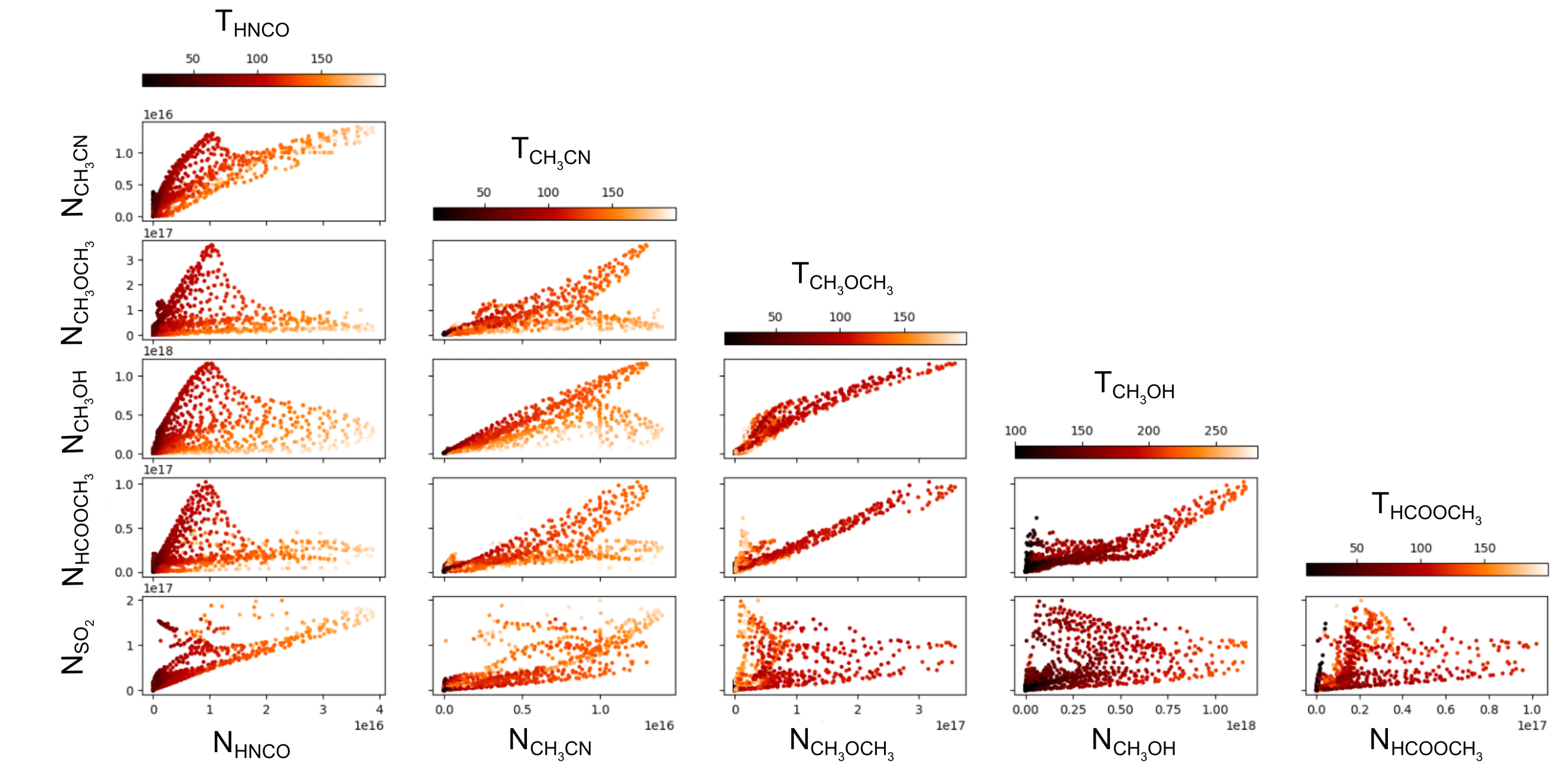}
  \caption{Plots comparing column densities in cm$^{-2}$ between pairs of molecules across the MM1 region. The color bar denotes temperature in K of the molecule on the x-axis.
  \label{fig:coldens_comparison}}
\end{figure*}

\begin{figure*}[ht]
  \includegraphics[width=\textwidth]{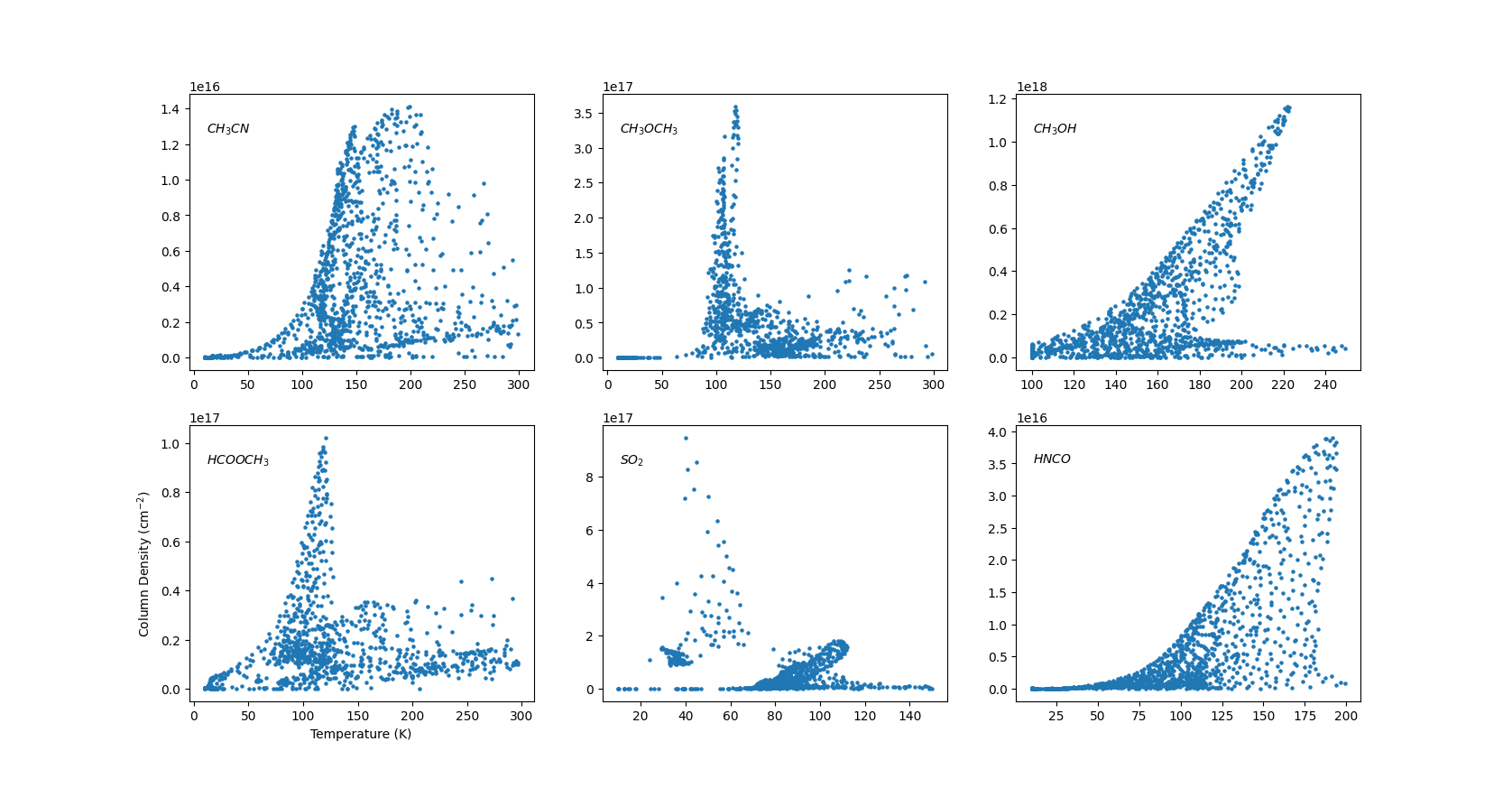}
  \caption{Plots comparing column density to temperature for each molecule across the MM1 region.
  \label{fig:coldensvstemp}}
\end{figure*}

Along with showing how the physical parameters of each molecule change across each region, it is also important to directly compare these parameters between each molecule. Figure \ref{fig:coldens_comparison} compares the calculated column density values of each individual pixel in the MM1 region. Generally, most of the plots show two distinct components based on whether the molecule peaks in MM1a or MM1b. For comparisons in which both molecules peak in either MM1a or MM1b, the relationship is almost linear. \ce{CH3OCH3}, \ce{CH3OH}, and \ce{HCOOCH3} directly exhibit this feature. \ce{SO2} correlates the least with the other molecules, most likely due to being the only sulfur molecule analyzed in this manner. Figure \ref{fig:coldensvstemp} directly compares column density and temperature for each molecule. Interestingly, each molecule apart from HNCO also individually exhibits two distinct components in line shape, albeit less obvious than seen in Figure \ref{fig:coldens_comparison}. For these molecules, one component has higher column density and a lower temperature, and the other component has lower column density and a higher temperature. This gives further evidence that different molecules are more abundant at either MM1a or MM1b.

\section{Discussion} \label{sec:image}

\begin{figure*}[ht]
  \includegraphics[width=0.9\textwidth]{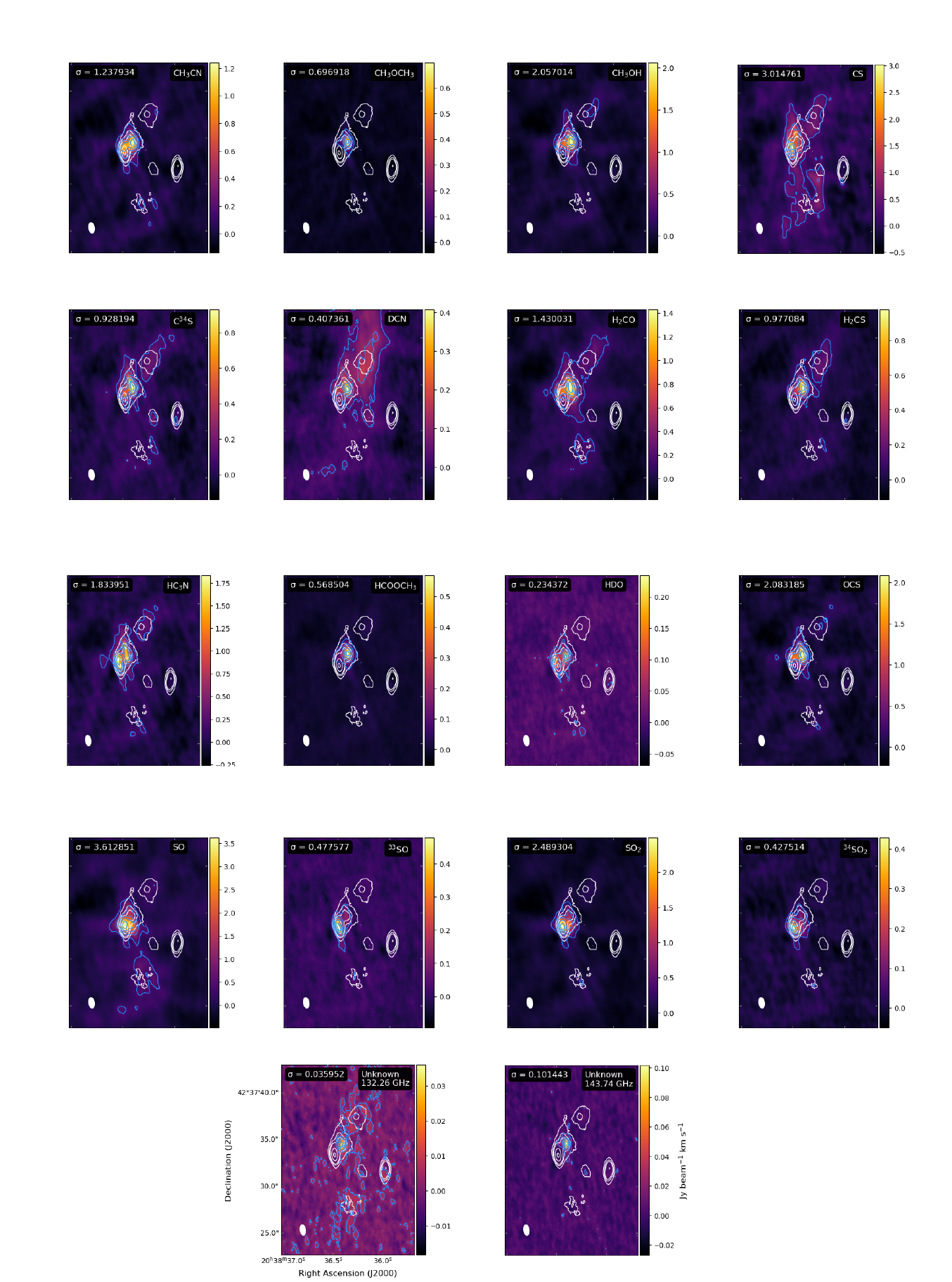}
  \caption{Integrated intensity maps of the MM1 region for all molecules seen in the high-resolution windows. For molecules with multiple transitions, the map was created by stacking all transitions together. Blue contour levels correspond to 0.1, 0.3, 0.5, 0.7, and 0.9 times $\sigma$ (reported in the top left corner). For all molecules, the $\sigma$ value signifies the maximum flux value reported. White contours correspond to the continuum levels at 1, 2, 3, 7, and 12 times $\sigma_{cont}$ ($\sigma_{cont}$ = 3.079 mJy beam$^{-1}$). Beam size is shown in the bottom left corner.
  \label{fig:mom0}}
\end{figure*}

\begin{figure*}[ht]
  \includegraphics[width=0.9\textwidth]{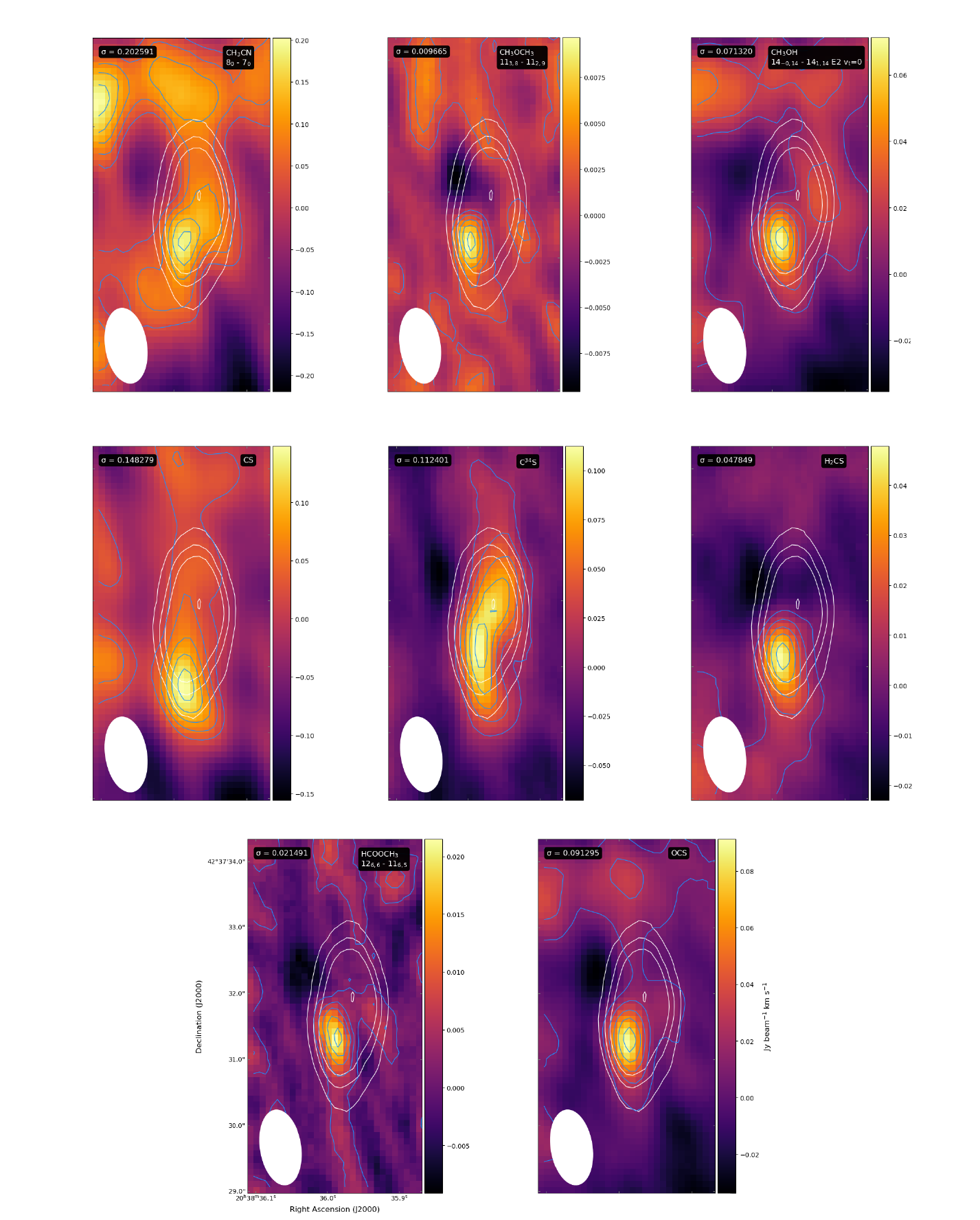}
  \caption{Integrated intensity maps of the MM2 region for molecules seen in the high-resolution windows. Blue contour levels correspond to 0.1, 0.3, 0.5, 0.7, and 0.9 times $\sigma$ (reported in the top left corner). For all molecules, the $\sigma$ value signifies the maximum flux value reported. White contours correspond to the continuum levels at 1, 2, 3, and 7 times $\sigma_{cont}$ ($\sigma_{cont}$ = 3.079 mJy beam$^{-1}$). Beam size is shown in the bottom left corner.
  \label{fig:MM2_mom0}}
\end{figure*}

\begin{figure*}[ht]
  \hspace{-50pt}\includegraphics[width=1.1\textwidth]{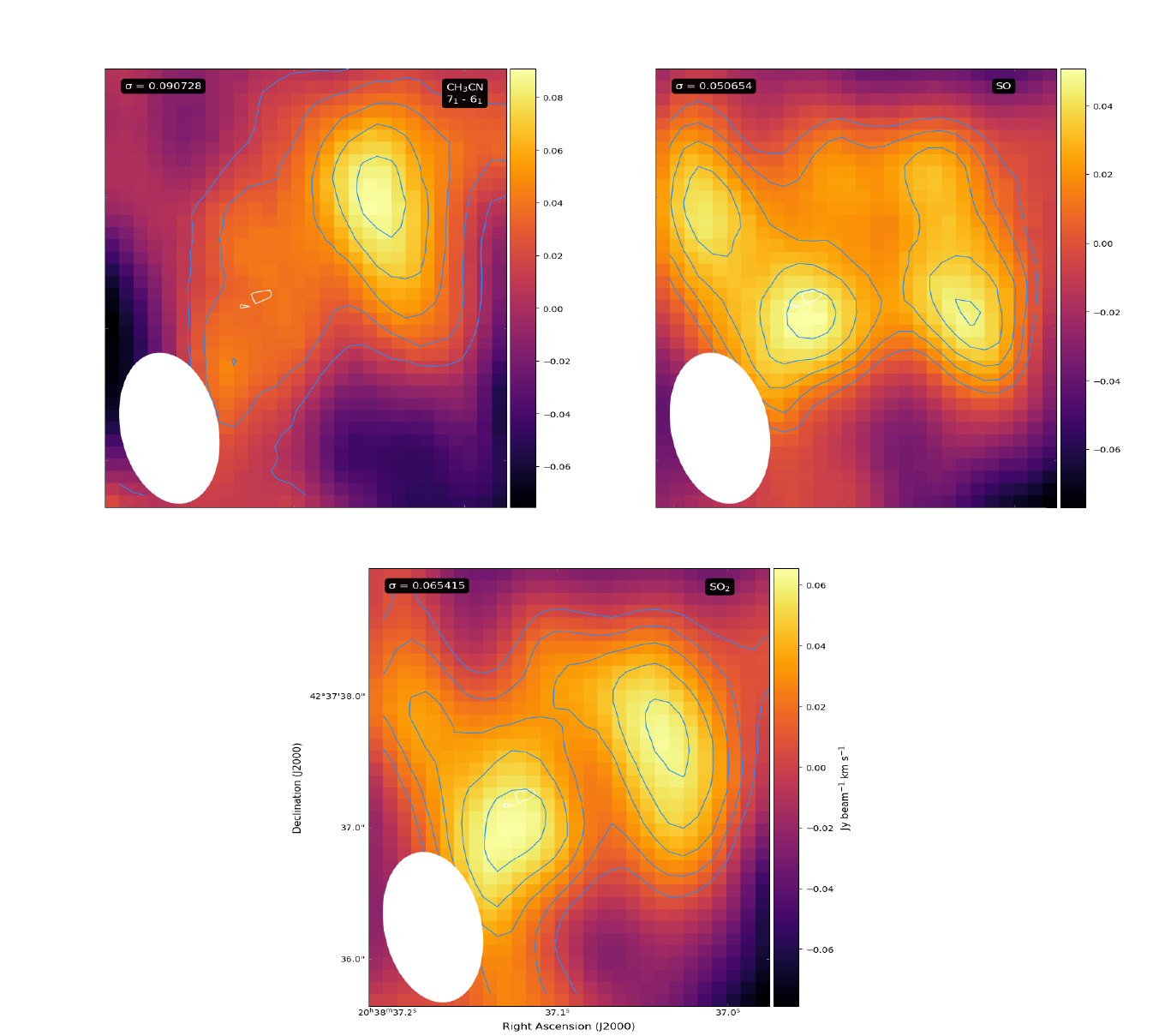}
  \caption{Integrated intensity maps of the MM3 region for molecules seen in the high-resolution windows. For molecules with multiple transitions, the map was created by stacking all transitions together. Blue contour levels correspond to 0.1, 0.3, 0.5, 0.7, and 0.9 times $\sigma$ (reported in the top left corner). For all molecules, the $\sigma$ value signifies the maximum flux value reported. White contours correspond to the continuum level at 1 times $\sigma_{cont}$ ($\sigma_{cont}$ = 3.079 mJy beam$^{-1}$). The map of \ce{SO2} was creating using the stacking method. Beam size is shown in the bottom left corner.
  \label{fig:MM3_mom0}}
\end{figure*}

\begin{figure*}[ht]
  \includegraphics[width=\textwidth]{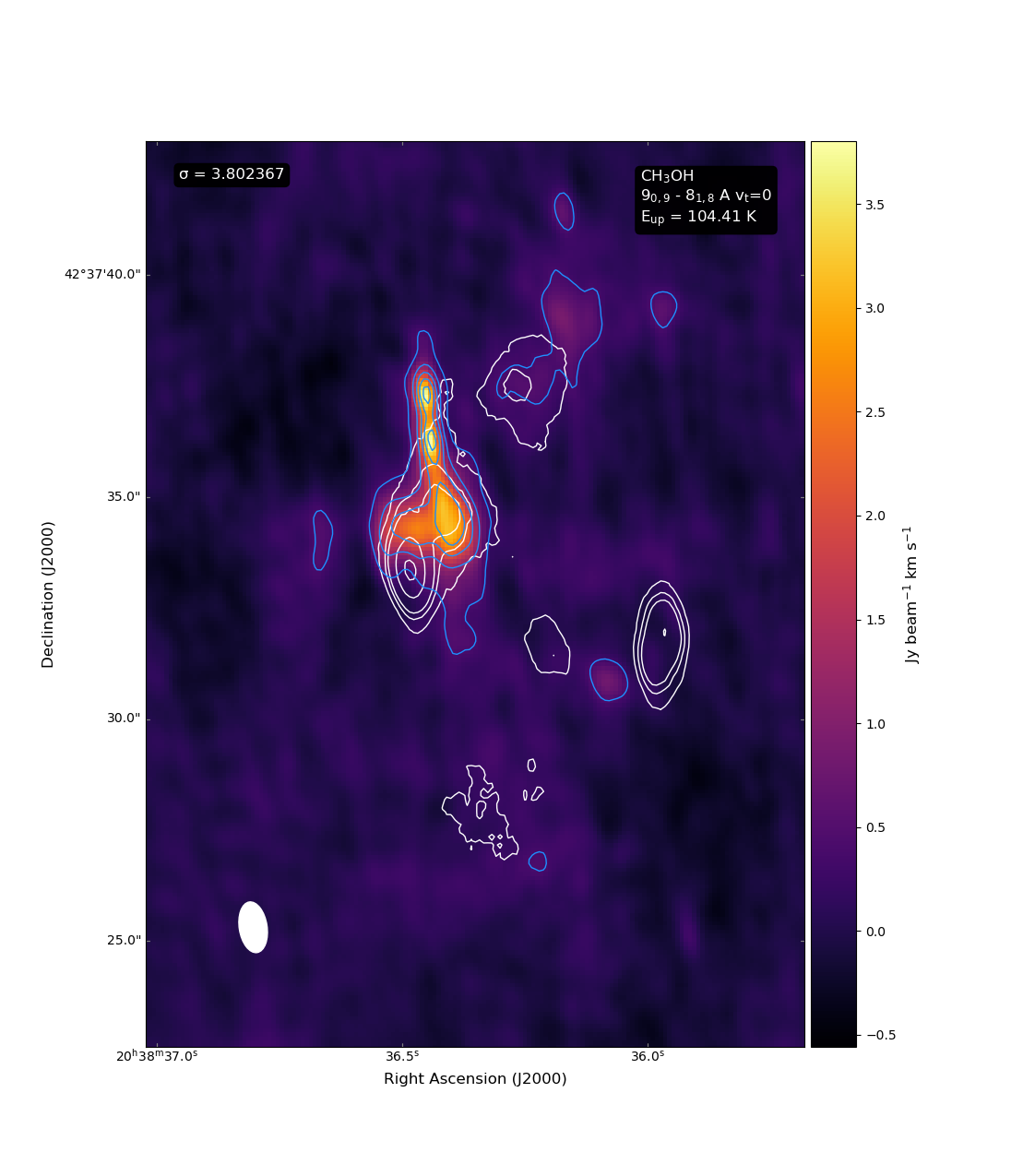}
  \caption{Integrated intensity map for the \ce{CH3OH} maser seen at 146.619 GHz. Blue contour levels correspond to 0.1, 0.3, 0.5, 0.7, and 0.9 times $\sigma$ (reported in the top left corner). The $\sigma$ value signifies the maximum flux value reported. White contours correspond to the continuum levels at 1, 2, 3, 7, and 12 times $\sigma_{cont}$ ($\sigma_{cont}$ = 3.079 mJy beam$^{-1}$). Beam size is shown in the bottom left corner.
  \label{fig:maser}}
\end{figure*}

From the 28 high-resolution windows, molecular emission from MM1, MM2, and MM3 were imaged via moment-0 maps using the Astropy\footnote{http://www.astropy.org} python library. These windows included numerous transitions for multiple molecules, as shown in Table \ref{tab:hires_lines}, with none of the reported transitions being affected by blending with other spectral lines. While this list describes every transition seen in the high-resolution datacubes, many more transitions and molecules were observed in the two low-resolution datacubes. For each transition seen in the high-resolution windows, emission was integrated over channels above 3$\sigma$, with the value of $\sigma$ changing dependent on the imaged cube (described in Appendix \ref{sec:windowspecs}). Stacked moment maps were made for molecules which had multiple transitions throughout all of the high-resolution windows by using a similar process to \cite{van_der_walt_protostellar_2021}. To do so, each transition was averaged together using a 1/$\sigma^2$ weighting scheme. Every molecule for which this stacking method was used had similar spatial distribution across transitions except for \ce{CH3OH}. Although every other methanol transition seen in the 28 high-resolution windows shows compact emission in MM1, the 9$_{0,9}$--8$_{1,8}$ A v$_t$=0 transition shows extended emission northward of MM1b with high intensity. This transition most likely traces the numerous methanol masers found north of MM1b shown in Figure 27 of \cite{fish_fullpolarization_2005}. While this transition is not a common indicator of a methanol maser, it has been predicted to be one at medium gas densities \citep{Nesterenok2021}. Due to having a vastly different spatial distribution compared to the other methanol transitions, this transition was not included in the stacking process and was instead imaged separately. Figure \ref{fig:mom0} shows the moment-0 map of each molecule seen in the high-resolution windows in the MM1 region, along with maps for two unidentified transitions. Due to the MM2 and MM3 regions having significantly lower fluxes than the MM1 region, separate moment-0 maps were created for molecules detected in these regions. These are shown in Figures \ref{fig:MM2_mom0} and \ref{fig:MM3_mom0}. For some molecules in the MM2 region, emission was only seen above 3$\sigma$ in specific transitions. Figure \ref{fig:maser} shows the MM1 moment-0 map of the methanol transition that traces the masers.

\begin{ThreePartTable}
\begin{TableNotes}
    \item[a] All transitions listed in this table were detected toward the MM1 region. However, not all listed transitions were also detected toward the MM2 and MM3 regions. This column denotes in which region the transition was detected above 3$\sigma$ in addition to MM1.
    \item[b] Peaks for CH$_3$OCH$_3$ are in the triplet form, representing the AE/EA, EE, and AA states, with the center peak of the triplet corresponding to the EE state. Therefore, for simplicity, the transitions reported here are that of the EE state.
    \item[c] The 3$_{2,1}$--2$_{2,0}$ E v$_t$=2 and 3$_{0,3}$--2$_{0,2}$ A v$_t$=2 transitions are within astronomical line widths of each other. Similarly, the 3$_{2,2}$--2$_{2,1}$ A v$_t$=1, 3$_{2,2}$--2$_{2,1}$ E v$_t$=1, and 3$_{2,1}$--2$_{2,0}$ A v$_t$=1 transitions are within astronomical line widths of each other. Due to severe blending, it is unclear which transition is represented in the spectrum.
\end{TableNotes}

\begin{center}
\begin{longtable}{ccccccc}
\caption{Molecular transitions covered in the high-resolution windows.}\\

\multicolumn{1}{c}{Molecule} & \multicolumn{1}{c}{Transition} & \multicolumn{1}{c}{Rest Frequency} & \multicolumn{1}{c}{E$_u$} & \multicolumn{1}{c}{SPW} & \multicolumn{1}{c}{MM2/MM3\tnote{a}} & \multicolumn{1}{c}{Weighted $\sigma$}\\
& & (GHz) & (K) & (mJy/beam)\\
\hline
\endhead

\midrule
\insertTableNotes
\endlastfoot

CH$_3$CN & 7$_{6}$--6$_{6}$ & 128.690 & 281.79 & 3 & & 4.531\\
& 7$_{5}$--6$_{5}$ & 128.717 & 203.28 & 3&\\
& 7$_{4}$--6$_{4}$ & 128.740 & 139.02 & 3&\\
& 7$_{3}$--6$_{3}$ & 128.757 & 89.02 & 3&\\
& 7$_{2}$--6$_{2}$ & 128.769 & 53.30 & 3&\\
& 7$_{1}$--6$_{1}$ & 128.777 & 31.87 & 3& MM3\\
& 8$_{6}$--7$_{6}$ & 147.073 & 288.84 & 25&\\
& 8$_{5}$--7$_{5}$ & 147.104 & 210.34 & 25&\\
& 8$_{4}$--7$_{4}$ & 147.129 & 146.08 & 25&\\
& 8$_{3}$--7$_{3}$ & 147.149 & 96.08 & 25&\\
& 8$_{2}$--7$_{2}$ & 147.163 & 60.36 & 25&\\
& 8$_{1}$--7$_{1}$ & 147.172 & 38.93 & 25&\\
& 8$_{0}$--7$_{0}$ & 147.175 & 31.79 & 25& MM2\\
\hline
CH$_3$OCH$_3$\tnote{b} & 11$_{3,8}$--11$_{2,9}$ & 133.268 & 72.87 & 9 & MM2 & 4.318\\
& 24$_{3,21}$--24$_{2,22}$ & 133.313 & 290.14 & 9&\\
& 12$_{2,11}$--12$_{1,12}$ & 135.267 & 76.28 & 14&\\
& 13$_{2,12}$--13$_{1,13}$ & 143.163 & 88.00 & 15&\\
& 6$_{3,3}$--6$_{2,4}$ & 144.859 & 31.77 & 19&\\
& 26$_{4,22}$--26$_{1,25}$ & 148.119 & 344.84 & 26&\\
& 9$_{3,7}$--9$_{2,8}$ & 149.570 & 53.64 & 27&\\
& 25$_{4,21}$--25$_{1,24}$ & 150.163 & 320.68 & 28&\\
& 21$_{2,19}$--21$_{1,20}$ & 150.594 & 220.15 & 28&\\
\hline
CH$_3$OH\tnote{c} & 12$_{1,11}$--11$_{2,10}$ A v$_t$=0 & 129.433 & 197.07 & 5 & & 4.432\\
& 6$_{2,5}$--7$_{1,6}$ A v$_t$=0 & 132.622 & 86.46 & 8&\\
& 6$_{1,6}$--5$_{- 0,5}$ E & 132.891 & 54.31 & 8&\\
& 20$_{- 4,16}$--19$_{- 5,14}$ E & 133.261 & 583.10 & 9&\\
& 5$_{2,3}$--6$_{1,6}$ E & 133.605 & 60.72 & 10&\\
& 12$_{3,9}$--13$_{2,11}$ E & 134.231 & 243.74 & 11&\\
& 8$_{2,7}$--7$_{3,4}$ A v$_t$=0 & 134.897 & 121.27 & 12&\\
& 7$_{- 3,5}$--8$_{- 2,7}$ E v$_t$=0 & 143.170 & 112.71 & 15&\\
& 3$_{2,1}$--2$_{2,0}$ E v$_t$=2 & 144.571 & 658.82 & 18&\\
& 3$_{0,3}$--2$_{0,2}$ A v$_t$=2 & 144.572 & 522.08 & 18&\\
& 3$_{- 2,2}$--2$_{- 2,1}$ E v$_t$=2 & 144.580 & 600.11 & 18&\\
& 3$_{1,3}$--2$_{1,2}$ E v$_t$=2 & 144.584 & 545.90 & 18&\\
& 3$_{1,3}$--2$_{1,2}$ A v$_t$=1 & 144.590 & 339.14 & 18&\\
& 3$_{2,2}$--2$_{2,1}$ A v$_t$=1 & 144.728 & 312.57 & 19&\\
& 3$_{2,2}$--2$_{2,1}$ E v$_t$=1 & 144.729 & 378.50 & 19&\\
& 3$_{2,1}$--2$_{2,0}$ A v$_t$=1 & 144.729 & 312.57 & 19&\\
& 3$_{-2,1}$--2$_{-2,0}$ E v$_t$=1 & 144.733 & 413.79 & 19&\\
& 3$_{1,3}$--2$_{1,2}$ v$_t$=1 & 144.735 & 305.37 & 19&\\
& 3$_{- 0,3}$--2$_{- 0,2}$ E v$_t$=1 & 144.736 & 314.46 & 19&\\
& 3$_{1,2}$--2$_{1,1}$ E v$_t$=1 & 144.750 & 427.27 & 19&\\
& 3$_{0,3}$--2$_{0,2}$ A v$_t$=1 & 144.768 & 437.54 & 19&\\
& 3$_{1,2}$--2$_{1,1}$ A v$_t$=1 & 144.879 & 339.16 & 19&\\
& 3$_{-0,3}$--2$_{-0,2}$ E v$_t$=0 & 145.093 & 27.05 & 20&\\
& 3$_{1,3}$--2$_{1,2}$ E v$_t$=0 & 145.097 & 19.50 & 20&\\
& 3$_{0,3}$--2$_{0,2}$ A v$_t$=0 & 145.103 & 13.93 & 20&\\
& 3$_{2,2}$--2$_{2,1}$ A v$_t$=0 & 145.124 & 51.64 & 20&\\
& 3$_{-2,2}$--2$_{-2,1}$ E v$_t$=0 & 145.126 & 36.17 & 20&\\
& 3$_{2,1}$--2$_{2,0}$ E v$_t$=0 & 145.126 & 39.83 & 20&\\
& 3$_{-1,2}$--2$_{-1,1}$ E v$_t$=0 & 145.132 & 34.97 & 20&\\
& 3$_{2,1}$--2$_{2,0}$ A v$_t$=0 & 145.133 & 51.64 & 20&\\
& 3$_{1,2}$--2$_{1,1}$ A v$_t$=0 & 146.368 & 28.59 & 23&\\
& 9$_{0,9}$--8$_{1,8}$ A v$_t$=0 & 146.619 & 104.41 & 24&\\
& 15$_{-0,15}$--15$_{1,15}$ E v$_t$=0 & 148.112 & 290.74 & 26&\\
& 14$_{-2,13}$--13$_{-3,11}$ E v$_t$=0 & 149.533 & 266.13 & 27&\\
& 14$_{-0,14}$--14$_{1,14}$ E2 v$_t$=0 & 150.142 & 256.14 & 28& MM2\\
\hline
CS & 3--2 & 146.969 & 14.11 & 25 & MM2&\\
\hline
C$^{34}$S & 3--2 & 144.617 & 11.80 & 18 & MM2&\\
\hline
DCN & 2--1 & 144.828 & 10.43 & 19&\\
\hline
H$_2$CO & 7$_{1,6}$--7$_{1,7}$ & 135.030 & 112.79 & 13 & & 4.185\\
& 2$_{0,2}$--1$_{0,1}$ & 145.603 & 10.48 & 21&\\
& 2$_{1,1}$--1$_{1,0}$ & 150.498 & 22.62 & 28&\\
\hline
H$_2$CS & 4$_{1,4}$--3$_{1,3}$ & 135.298 & 29.40 & 14 & MM2&\\
\hline
HC$_{3}$N & 16--15 & 145.561 & 59.38 & 21&\\
\hline
HCOOCH$_3$ & 12$_{0,12}$--11$_{0,11}$ & 132.245 & 42.43 & 7 & & 4.021\\
& 11$_{1,10}$--10$_{1,9}$ E & 132.922 & 40.39 & 8&\\
& 11$_{1,10}$--10$_{1,9}$ A & 132.929 & 40.38 & 8&\\
& 11$_{7,4}$--10$_{7,3}$ & 135.290 & 71.48 & 14&\\
& 12$_{6,6}$--11$_{6,5}$ & 148.046 & 69.96 & 26 & MM2&\\
& 12$_{4,8}$--11$_{4,7}$ E & 150.601 & 57.04 & 28&\\
& 12$_{4,8}$--11$_{4,7}$ A & 150.618 & 57.02 & 28&\\
\hline
HDO & 4$_{2,2}$--4$_{2,3}$ & 143.727 & 319.17 & 17&\\
\hline
OCS & 12--11 & 145.947 & 45.53 & 22 & MM2&\\
\hline
SiO & 3--2 & 130.269 & 12.50 & 6&\\
\hline
SO & 3$_{3}$--2$_{2}$ & 129.139 & 25.51 & 4 & MM3&\\
\hline
$^{33}$SO & 3$_3$--2$_2$ & 127.859 & 25.41 & 1&\\
\hline
SO$_2$ & 12$_{2,10}$--12$_{1,11}$ & 128.605 & 82.58 & 3 & MM3 & 4.438\\
& 12$_{1,11}$--11$_{2,10}$ & 129.106 & 76.41 & 4 & MM3&\\
& 10$_{2,8}$--10$_{1,9}$ & 129.515 & 60.93 & 5 & MM3&\\
& 14$_{2,12}$--14$_{1,13}$ & 132.745 & 108.12 & 8 & MM3&\\
& 10$_{4,6}$--11$_{3,9}$ & 146.550 & 89.83 & 24 & MM3&\\
& 4$_{2,2}$--4$_{1,3}$ & 146.606 & 19.03 & 24 & MM3&\\
\hline
$^{34}$SO$_2$ & 8$_{2,6}$--8$_{1,7}$ & 128.669 & 42.83 & 3&\\
\hline
Unknown Line & - & 132.258 & - & 7&\\
& - & 143.738 & - & 17&\\

\end{longtable}
\end{center}
\label{tab:hires_lines}
\end{ThreePartTable}

\begin{figure*}[ht]
  \includegraphics[width=0.5\textwidth]{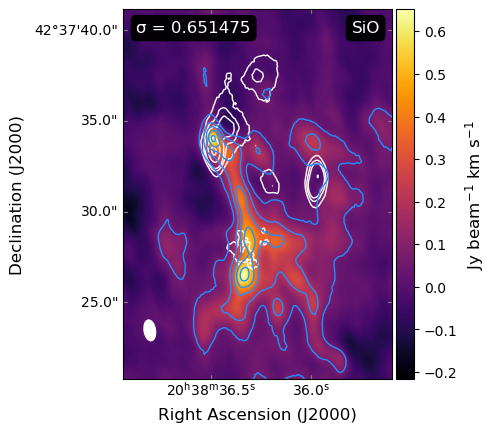}
  \includegraphics[width=0.5\textwidth]{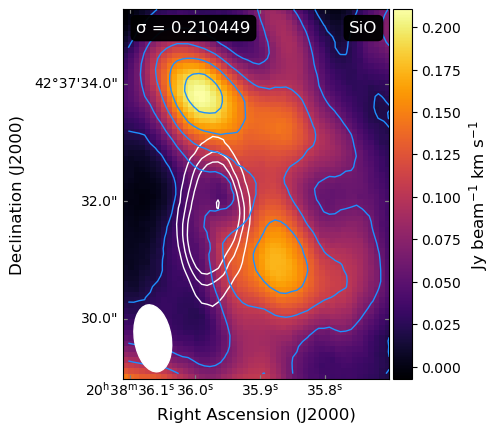}
  \caption{Integrated intensity maps for SiO in MM1 (left) and MM2 (right). Blue contour levels correspond to 0.1, 0.3, 0.5, 0.7, and 0.9 times $\sigma$ (reported in the top left corner). The $\sigma$ value signifies the maximum flux value reported. White contours correspond to the continuum levels at 1, 2, 3, 7, and 12 times $\sigma_{cont}$ ($\sigma_{cont}$ = 3.079 mJy beam$^{-1}$). Beam size is shown in the bottom left corner.
  \label{fig:SiO}}
\end{figure*}

As mentioned above, SiO has been known to trace molecular outflows in star-forming regions. Specifically, \cite{shepherd_nature_2004} found that the molecular emission from the \textit{J} = 2--1 transition extends $\sim$ 20\arcsec southward from the MM1 region. Figure \ref{fig:SiO}, the moment-0 maps for SiO in MM1 and MM2, show similar features to this result. Interestingly, \cite{Minh2010} and \cite{carrasco-gonzalez_bright_2010} saw much more compact emission from the \textit{J} = 5--4, \textit{J} = 8--7, and \textit{J} = 1--0  transitions. While both previous results and those presented here show emission peaking around MM1a, there is no direct correlation between SiO emission and previously reported outflows in W75N \citep{shepherd_clustered_2003,shepherd_nature_2004}. \cite{carrasco-gonzalez_bright_2010} instead proposes that the SiO emission traces a new outflow associated with VLA3.

Previous studies have shown that molecules can generally be grouped into two categories depending on whether their emission peaks in MM1a or MM1b, with the complex organics peaking in MM1b and sulfur-bearing molecules peaking in MM1a \citep{Minh2010, van_der_walt_protostellar_2021}. The moment-0 maps presented here support this argument. CS, SiO, SO, $^{33}$SO, \ce{SO2}, and $^{34}$SO$_2$ peak in MM1a. DCN, \ce{CH3OCH3}, \ce{CH3OH}, C$^{34}$S, \ce{H2CO}, \ce{HCOOCH3}, \ce{H2CS}, OCS, and the two unknown lines peak in MM1b. A third category can be used to describe \ce{CH3CN}, HDO, and \ce{HC3N} $v_7$, as all three show strong emission in both MM1a and MM1b. These results concur with \cite{Minh2010} in that there is a hot core located at MM1b, as complex organic molecules are often tracers of similar sources (e.g. \cite{Blake1998}).

For the first time, molecular emission was directly imaged in the MM2 and MM3 region. Every molecule seen in MM2, apart from \ce{C^34S}, shows compact emission on the southern portion of the continuum core. \ce{C^34S}, while peaking in a similar location to the other molecules, has emission extending northward. In MM3, both SO and \ce{SO2} show two distinct emission peaks. This could be due to the presence of two young stellar objects, as was theorized by \cite{Shepherd2001}. The presence of OCS in the spectrum also supports this, as both OCS and SO tend to trace high temperature chemistry in early hot cores \citep{Wakelam2004,Tycho2021}. All molecules in both MM2 and MM3 only show emission at a level roughly an order of magnitude lower than that in MM1.

%Furthermore, the maps show that molecules can be seen with either compact or extended emission relative to the MM1 continuum contours. Molecules with compact emission include \ce{CH3CN}, \ce{CH3OCH3}, \ce{HCOOCH3}, HDO, $^{33}$SO, $^{34}$SO$_2$, and the two unknown transitions. Molecules with extended emission include DCN, \ce{CH3OH}, CS, C$^{34}$S, \ce{H2CO}, \ce{HC3N} $v_7$, \ce{H2CS}, OCS, SO, and \ce{SO2}.

\begin{figure*}[ht]
  \hspace{-90pt}\includegraphics[width=1.5\textwidth]{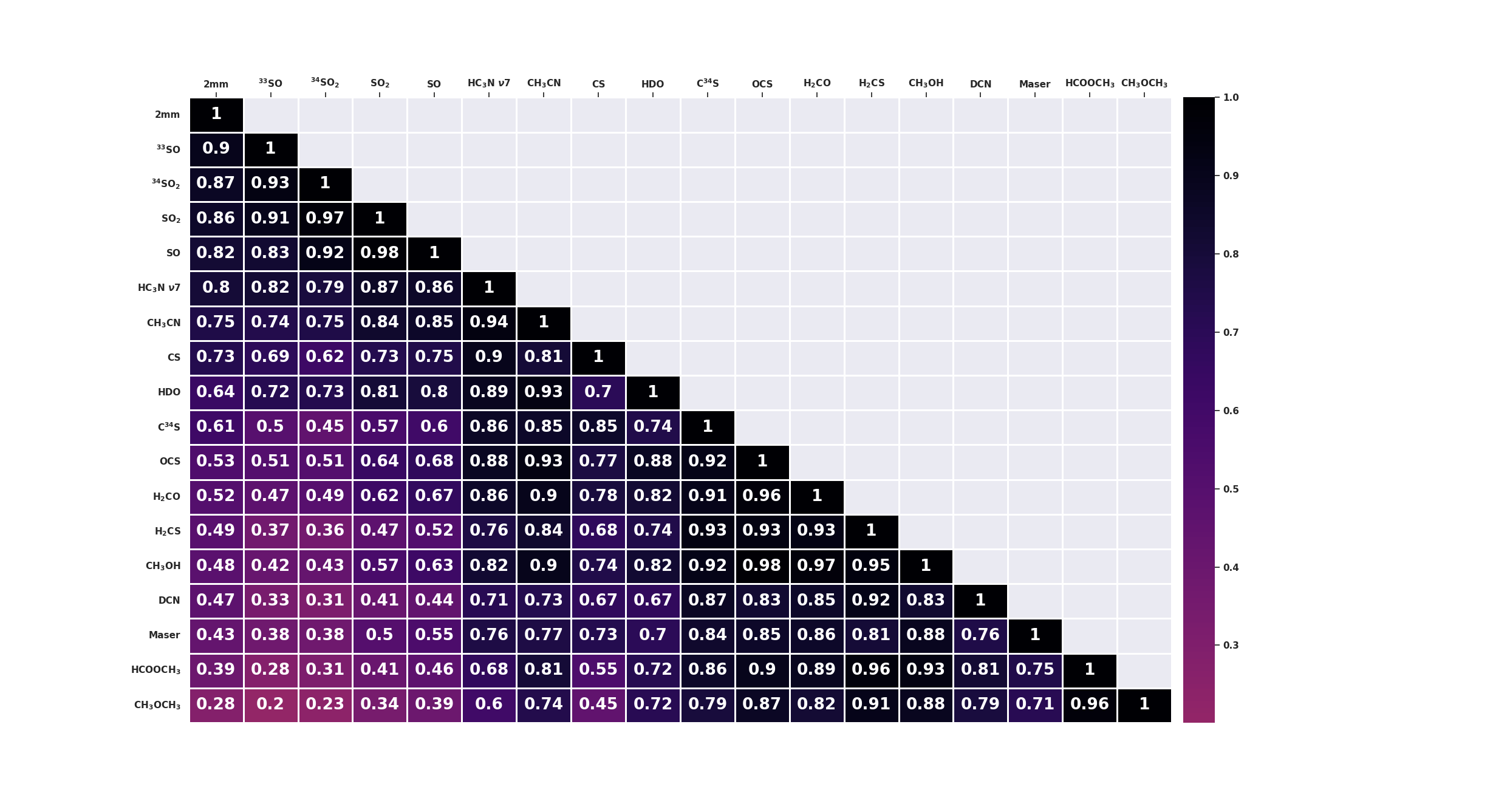}
  \caption{Correlation matrix of the MM1 region using the integrated intensity maps shown in Figure \ref{fig:mom0}, the 130 GHz continuum labeled as "2mm", and the integrated intensity map of the \ce{CH3OH} maser-tracing transition at 146.619 GHz labeled as "Maser".
  \label{fig:corrm1}}
\end{figure*}

\begin{figure*}[ht]
  \hspace{-90pt}\includegraphics[width=1.5\textwidth]{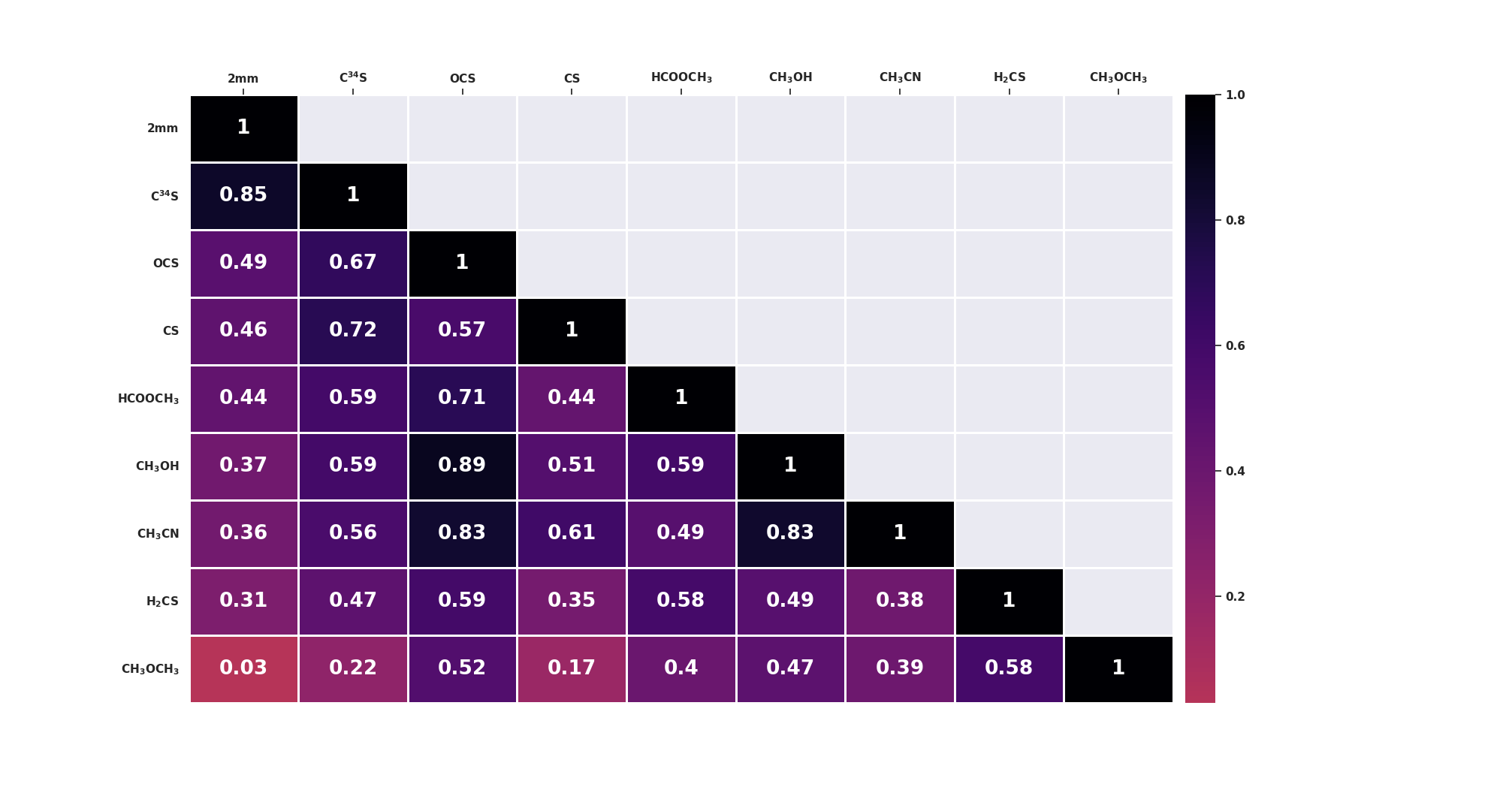}
  \caption{Correlation matrix of the MM2 region using the integrated intensity maps shown in Figure \ref{fig:MM2_mom0} and the 130 GHz continuum labeled as "2mm".
  \label{fig:corrm2}}
\end{figure*}

\begin{figure*}[ht]
  \includegraphics[width=\textwidth]{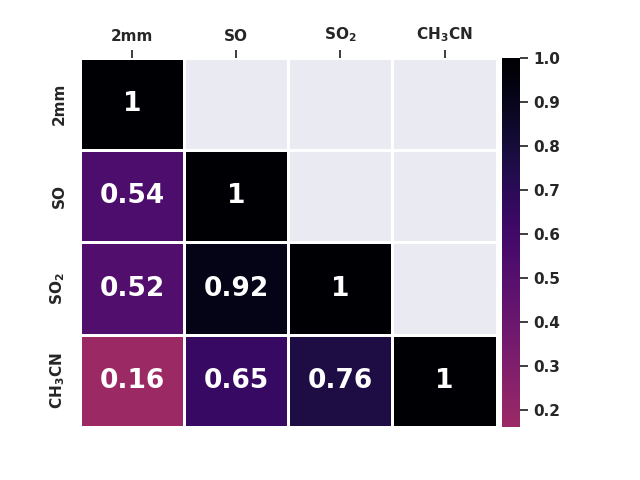}
  \caption{Correlation matrix of the MM3 region using the integrated intensity maps shown in Figure \ref{fig:MM3_mom0} and the 130 GHz continuum labeled as "2mm".
  \label{fig:corrm3}}
\end{figure*}

Figures \ref{fig:corrm1}, \ref{fig:corrm2}, and \ref{fig:corrm3} show the calculated Pearson cross-correlation coefficient between each pair of moment-0 maps and the continuum for each region. This coefficient measures the linear correlation of molecular emission between molecules on a spatial scale, with the emission being more similar as the coefficient goes to a value of one. This method has been used in multiple other works to study the chemistry of star-forming regions on a spatial scale \citep{Guzman2018, Law2021, Thompson2023}. However, it is also important to note that a high correlation coefficient does not intrinsically imply a chemical link between molecules. To calculate this value for the MM1 region, a mask was defined that encompassed a right ascension of 20$^{h}$38$^{m}$36.3$^{s}$ -- 36.7$^{s}$ and a declination of 42$^{o}$37\arcmin38\arcsec.2 -- 31\arcsec.9 in order to account for both MM1a and MM1b. For the MM2 and MM3 regions, this mask was defined with the same boundaries as their respective moment-0 maps. The cross-correlation coefficient was calculated using the pixels within these masks. For the MM1 region, in addition to the maps shown in Figure \ref{fig:mom0} and the continuum, molecules were also correlated with the moment-0 map of the \ce{CH3OH} maser-tracing transition. SiO was not included in this analysis for any region due to its high-intensity extended features not seen in any other molecule. Due to the low total flux in both the MM2 and MM3 regions, creating a correlation matrix for molecules proved to be difficult. At such levels, emission from outside the continuum contours has a much larger affect on the correlation. While further observations on the MM2 and MM3 regions are needed to understand the spatial complexity of present molecules, an initial analysis was performed here.

As expected, the pairs of molecules that correlate well across MM1 are those that peak in either MM1a or MM1b. Most of the sulfur-bearing molecules, which peak in MM1a, have strong correlations with one another and the continuum. All of the organic molecules, which peak in MM1b, have strong correlations with one another and weak correlations with the continuum. This spatial differentiation between the organics and sulfur-bearing molecules implies that MM1a and MM1b are at different stages of evolution, with MM1b being the younger of the two \citep{van_der_walt_protostellar_2021}. Organic molecules in hot cores can be formed from  ultraviolet radiation photoprocessing and thermal processing of ices \citep{MunozCaro2002, Herbst2009, Oberg2011, Altwegg2019}. Conversely, the abundance of sulfur-bearing molecules in MM1a is potentially indicative of hot gas-phase chemistry due to shocks. Shocks have been seen to affect sulfuric chemistry in numerous other sources via multiple reactions, including the enhancement of SO, the increase of S$^{+}$, and the release of sulfur from \ce{H2S} \citep{Charnley1997, Minh2010, Guzman2018, Thompson2023}. These shocks are also evident due to the extended SiO emission, as SiO is known to trace shocked regions of gas \citep{hartquist_molecular_1980, Minh2010}.

Most molecules in MM2 have relatively low correlation values when compared to those seen in MM1. The three molecules that give the highest correlation are \ce{CH3CN}, \ce{CH3OH}, and OCS, which also are correlated highly in MM1. \ce{CH3OCH3} is least correlated with the other molecules, likely due to having emission at the lowest flux values. In MM3, SO and \ce{SO2} have a much higher correlation value with each other as opposed to with \ce{CH3CN}. This is expected as both are sulfur-bearing molecules.

The results for W75N presented here are part of a NOEMA survey to follow-up on the single-dish observations of \cite{WidicusWeaver2017}.  Previously, \cite{Thompson2023} and \cite{giese_mapping_2023} analyzed the molecular composition of the neighboring cores W3(\ce{H2O}) and W3(OH) in the W3 high-mass star-forming region. In the analysis of W3, six molecules accounted for the majority of lines seen across both cores: \ce{CH3CH2CN}, \ce{CH3CN}, \ce{CH3OCH3}, \ce{CH3OH}, \ce{HCOOCH3}, and \ce{SO2}. Of these molecules, all but \ce{CH3CH2CN} are also abundant in the MM1 core of W75N, allowing for direct comparison of column densities and temperatures between the molecules in these two sources. \ce{CH3CN}, \ce{CH3OH}, and \ce{SO2} all have higher column densities in MM1 than in W3, while the column densities of \ce{CH3OCH3} and \ce{HCOOCH3} were comparable. Further abundance comparisons between sources will be performed in a future analysis once more sources are studied. Generally, all molecules showed similar temperatures between the two sources. However, \ce{HCOOCH3} was much hotter in W3 with temperatures peaking $\sim$160 K, in contrast with the derived temperature of $\sim$90 K in the continuum peak of MM1b. 

While the comparisons between sources are useful, an important difference must be acknowledged when making direct comparisons between the two W3 cores and between the MM1a and MM1b regions. In \cite{Thompson2023}, the analysis showed that although W3(\ce{H2O}) and W3(OH) had distinct chemical inventories and molecular properties, the two cores generally showed similar results. However, the two sub-regions MM1a and MM1b instead have vastly different chemical inventory and derived parameters. This demonstrates that each star forming region may have unique chemical mechanisms at play, the physical origins of which have yet to be determined.  Our larger survey project that combines single-dish and interferometric observations of all 30 sources examined by \citep{WidicusWeaver2017} will hopefully reveal what might be driving these differences.  Further results from other sources in our ongoing NOEMA observations will allow for a more comprehensive comparative analysis.

\section{Summary}

We have observed the W75N star-forming region using the IRAM/NOEMA interferometer and analyzed the chemical composition and distribution in the MM1, MM2, and MM3 continuum cores. The broadband spectra extracted toward each core were analyzed with GOBASIC to identify over twenty molecules across the entire region, with the first chemical analysis of this kind being done on MM2 and MM3. COMs were detected in these two cores for the first time. Column density, rotational temperature, and velocity maps were created to spatially characterize six different molecules across the entire region. From these maps, we find a high correlation in the column densities of \ce{CH3OCH3}, \ce{CH3OH}, and \ce{HCOOCH3}. The results for MM1a and MM1b generally agree with previous works on W75N \citep{Minh2010, WidicusWeaver2017, van_der_walt_protostellar_2021}. Like these previous works, we also find a distinct difference in chemical composition between MM1a and MM1b, with most sulfur-bearing molecules tracing MM1a and most COMs tracing MM1b. Additionally, we created moment-0 maps for seventeen molecules, two unidentified lines, and a maser-tracing methanol transition found in the 28 high-resolution windows. This includes the first imaging of molecular emission in MM2 and MM3. By comparing all of the moment-0 maps in the MM1 region using a Pearson correlation coefficient analysis, the results further support the differentiation between MM1a and MM1b. We find that MM1b likely hosts a hot core with rich COM chemistry, whereas MM1a is affected by shocks, causing SiO emission to trace southward from the region. {We also performed an initial correlation analysis on the MM2 and MM3 regions, although further observations are necessary in order to better understand the spatial complexity of these cores.} Finally, when comparing results to those of the W3 star-forming region given by \cite{Thompson2023} and \cite{giese_mapping_2023}, we find similarity in which molecules were detected, although with differing physical parameters and distribution trends.  These results indicate that there is a physical mechanism driving chemical differentiation in star-forming cores, the origin of which has yet to be determined.

\section{Acknowledgements}

S.L.W.W., M.M.G., and W.E.T. thank the University of
Wisconsin–Madison for S.L.W.W.'s startup support and access
to NOEMA time that enabled this research. We thank Orsolya Feher from IRAM for support in setting up the observations and initial data reduction. Part of this research was carried out
at the Jet Propulsion Laboratory, California Institute of
Technology, under a contract with the National Aeronautics
and Space Administration (80NM0018D0004).

\software{Astropy \citep{astropy:2013, astropy:2018, astropy:2022}, pybaselines \citep{pybaselines}}

\begin{appendix}

\section{Spectral Window Summary} \label{sec:windowspecs}

A summary of the observational setup is shown in Table \ref{tab:window}. The frequency range of all 28 spectral windows is reported along with the RMS value of each cube. This RMS value was used when creating the moment maps as described above.

\begin{table*}
  \centering
  
    \caption{Summary of High-Resolution Spectral Windows}
    \label{tab:window}

    \begin{tabular}{c|c|c} 
        \hline
        \hline
        Spectral Window & Frequency Range (GHz) & RMS (mJy beam$^{-1}$)\\
        \hline
        1    &  127.823 - 127.888   & 4.62\\

        2    &  127.950 - 128.080   & 4.96\\

        3    &  128.589 - 128.849   & 4.52\\

        4    &  129.103 - 129.168   & 4.40\\

        5    &  129.354 - 129.746   & 4.22\\

        6    &  130.190 - 130.319   & 3.08\\

        7    &  132.239 - 132.303   & 3.34\\

        8    &  132.557 - 132.944   & 3.90\\

        9    &  133.133 - 133.328   & 3.79\\

        10    &  133.382 - 133.647   & 3.84\\
        
        11    &  134.157 - 134.287   & 3.79\\

        12    &  134.861 - 134.926   & 3.90\\

        13    &  134.989 - 135.055   & 3.92\\

        14    &  135.245 - 135.311   & 4.06\\

        15    &  143.116 - 143.181   & 3.87\\

        16    &  143.435 - 143.566   & 4.10\\

        17    &  143.692 - 143.756   & 3.91\\

        18    &  144.522 - 144.653   & 4.37\\

        19    &  144.714 - 144.910   & 4.75\\

        20    &  145.036 - 145.163   & 4.55\\

        21    &  145.547 - 145.612   & 4.45\\

        22    &  145.931 - 145.996   & 4.43\\
    
        23    &  146.315 - 146.379   & 4.39\\
    
        24    &  146.507 - 146.635   & 4.42\\
    
        25    &  146.954 - 147.213   & 4.54\\
    
        26    &  148.042 - 148.172   & 4.66\\
    
        27    &  149.514 - 149.579   & 5.32\\
    
        28    &  150.091 - 150.666   & 5.64\\
        \hline
    \end{tabular}

\end{table*}

\section{Parameter Map Figures} \label{sec:maps}

This section provides the parameter maps for the five molecules that fit in both MM1 and either MM2 or MM3. The complete figure set (5 images) is available in the online journal.

\figsetstart
\figsetnum{15}
\figsettitle{Parameter Maps}
\figsetgrpstart
\figsetgrpnum{15.1}
\figsetgrptitle{CH$_3$OCH$_3$ Maps}
\figsetplot{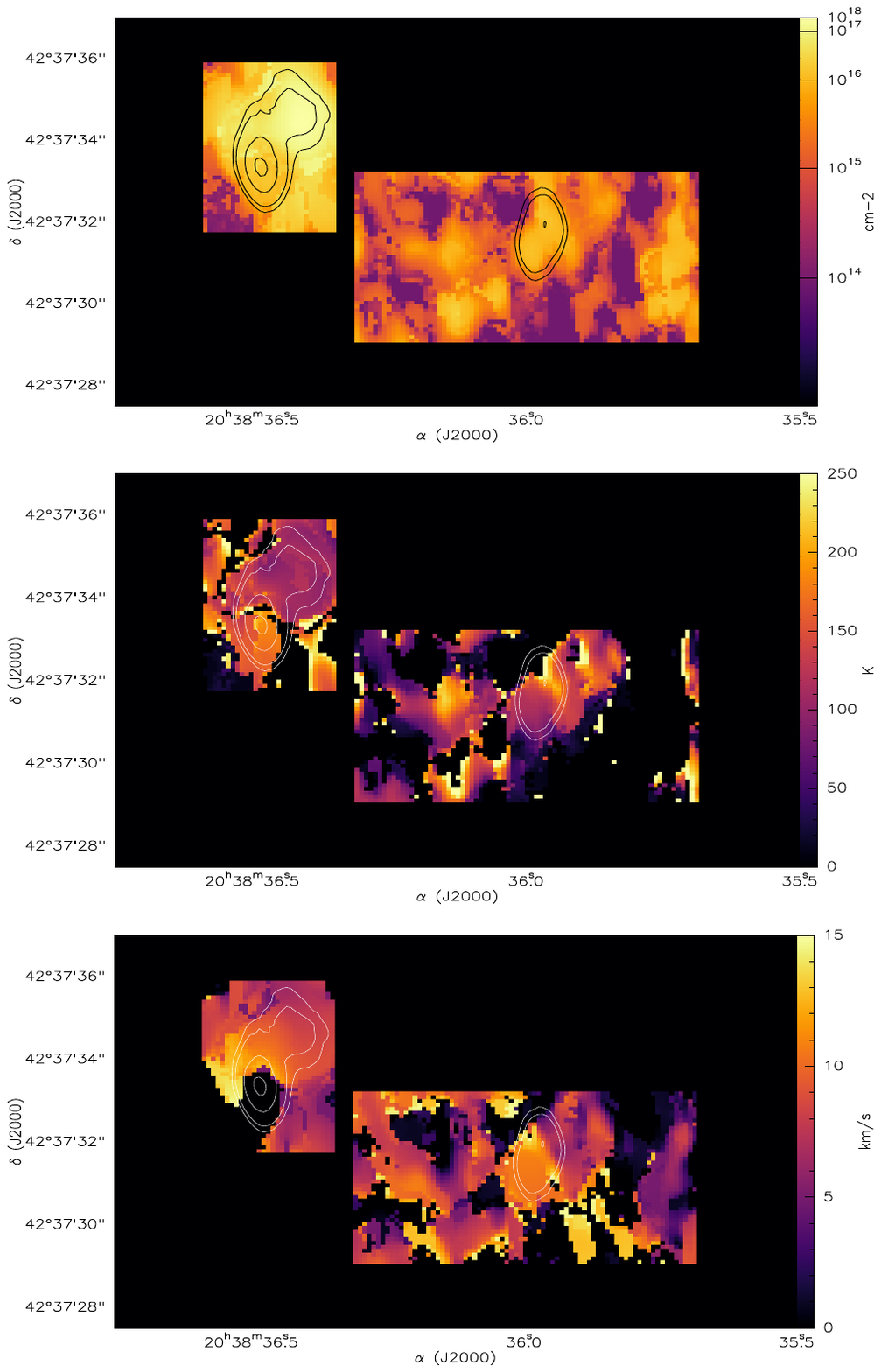}
\figsetgrpnote{Parameter maps for \ce{CH3OCH3} across MM1 and MM2 (from top to bottom: column density, kinetic temperature, and v$_{lsr}$). Black and white contours correspond to the continuum levels at 2, 3, 7, and 12 times $\sigma$ ($\sigma$ = 3.079 mJy beam$^{-1}$). The colors of the contours are arbitrary and differ between maps purely for aesthetic purposes. In the MM2 region, most pixels have uncertainties greater than the derived parameter values. The analysis of parameters across this region using this mapping method should therefore be seen as an initial estimate rather than a quantitative result.}
\figsetgrpend

\figsetgrpstart
\figsetgrpnum{15.2}
\figsetgrptitle{CH$_3$OH Maps}
\figsetplot{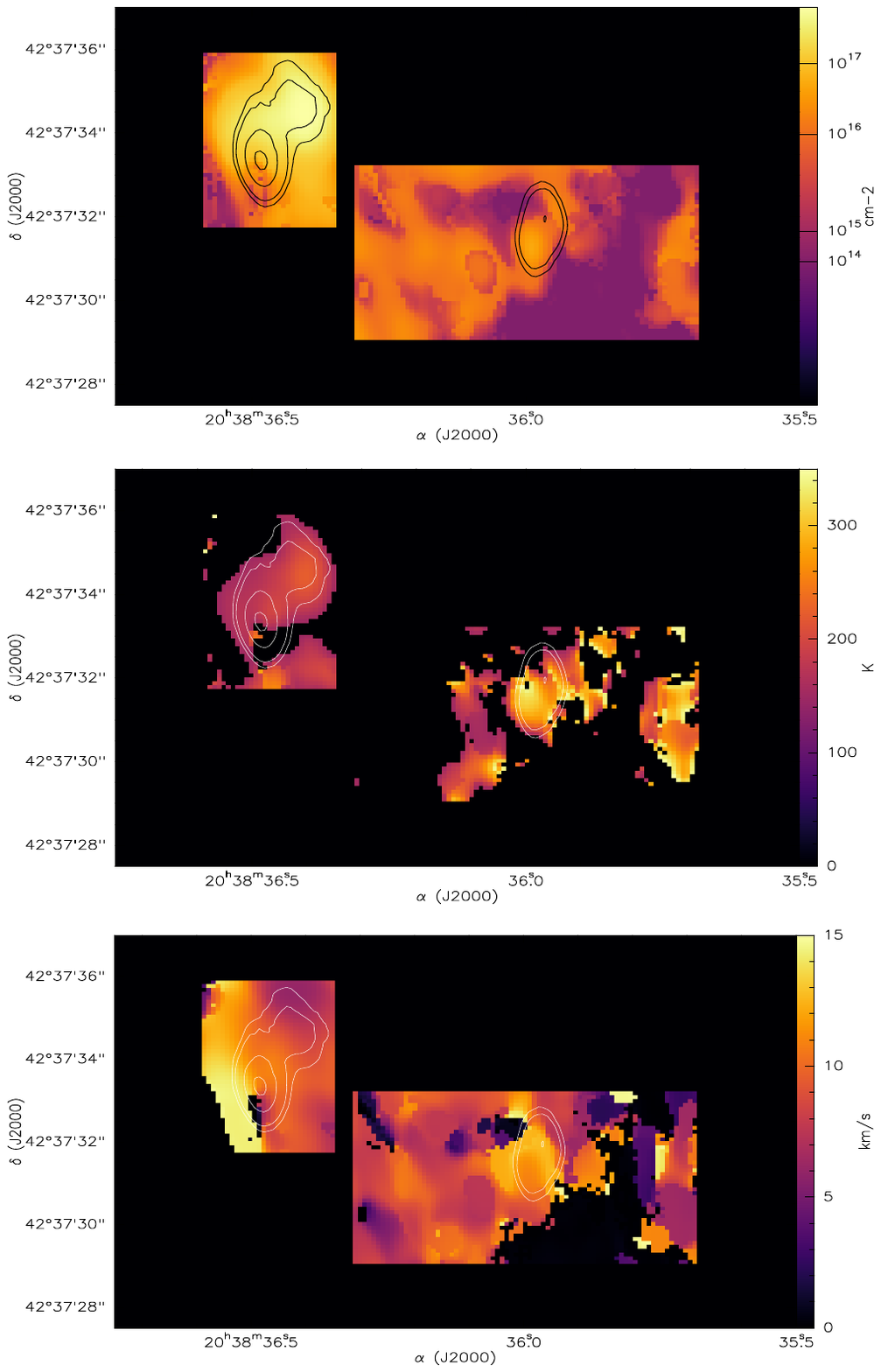}
\figsetgrpnote{Parameter maps for \ce{CH3OH} across MM1 and MM2 (from top to bottom: column density, kinetic temperature, and v$_{lsr}$). Black and white contours correspond to the continuum levels at 2, 3, 7, and 12 times $\sigma$ ($\sigma$ = 3.079 mJy beam$^{-1}$). The colors of the contours are arbitrary and differ between maps purely for aesthetic purposes. In the MM2 region, most pixels have uncertainties greater than the derived parameter values. The analysis of parameters across this region using this mapping method should therefore be seen as an initial estimate rather than a quantitative result.}
\figsetgrpend

\figsetgrpstart
\figsetgrpnum{15.3}
\figsetgrptitle{HCOOCH$_3$ Maps}
\figsetplot{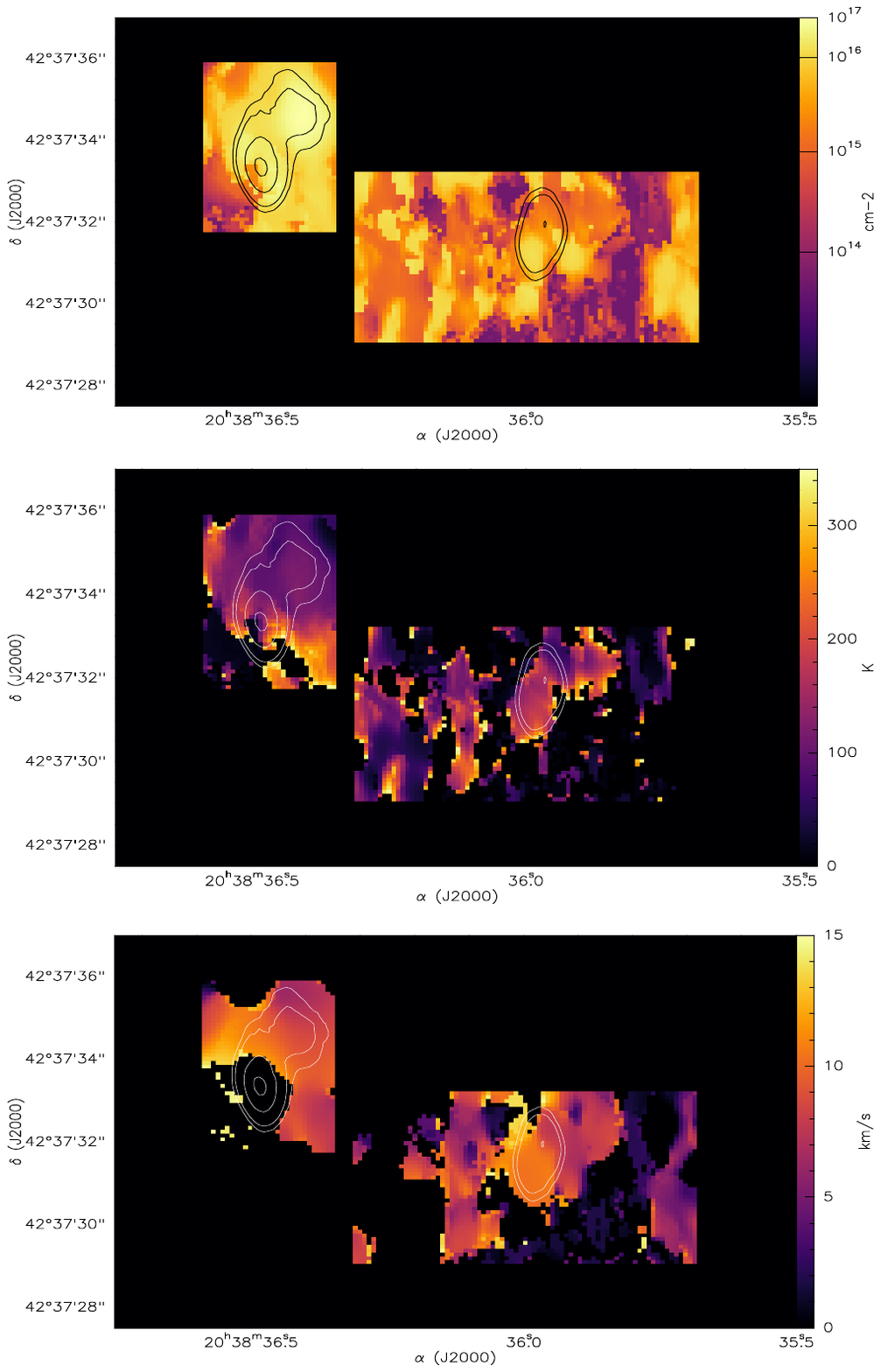}
\figsetgrpnote{Parameter maps for \ce{HCOOCH3} across MM1 and MM2 (from top to bottom: column density, kinetic temperature, and v$_{lsr}$). Black and white contours correspond to the continuum levels at 2, 3, 7, and 12 times $\sigma$ ($\sigma$ = 3.079 mJy beam$^{-1}$). The colors of the contours are arbitrary and differ between maps purely for aesthetic purposes. In the MM2 region, most pixels have uncertainties greater than the derived parameter values. The analysis of parameters across this region using this mapping method should therefore be seen as an initial estimate rather than a quantitative result.}
\figsetgrpend

\figsetgrpstart
\figsetgrpnum{15.4}
\figsetgrptitle{HNCO Maps}
\figsetplot{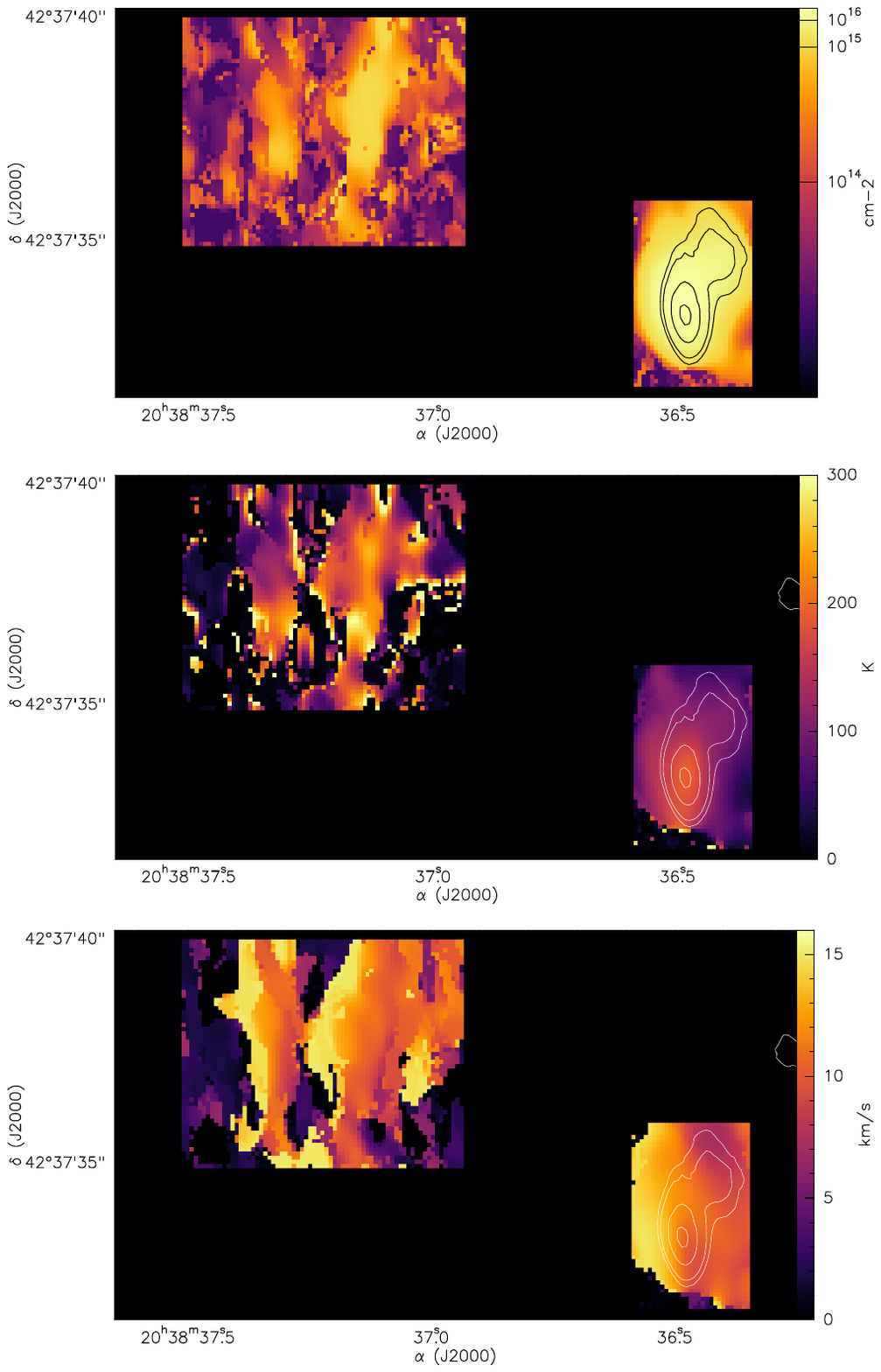}
\figsetgrpnote{Parameter maps for HNCO across MM1 and MM3 (from top to bottom: column density, kinetic temperature, and v$_{lsr}$). Black and white contours correspond to the continuum levels at 2, 3, 7, and 12 times $\sigma$ ($\sigma$ = 3.079 mJy beam$^{-1}$). The colors of the contours are arbitrary and differ between maps purely for aesthetic purposes. In the MM3 region, most pixels have uncertainties greater than the derived parameter values. The analysis of parameters across this region using this mapping method should therefore be seen as an initial estimate rather than a quantitative result.}
\figsetgrpend

\figsetgrpstart
\figsetgrpnum{15.5}
\figsetgrptitle{SO$_2$ Maps}
\figsetplot{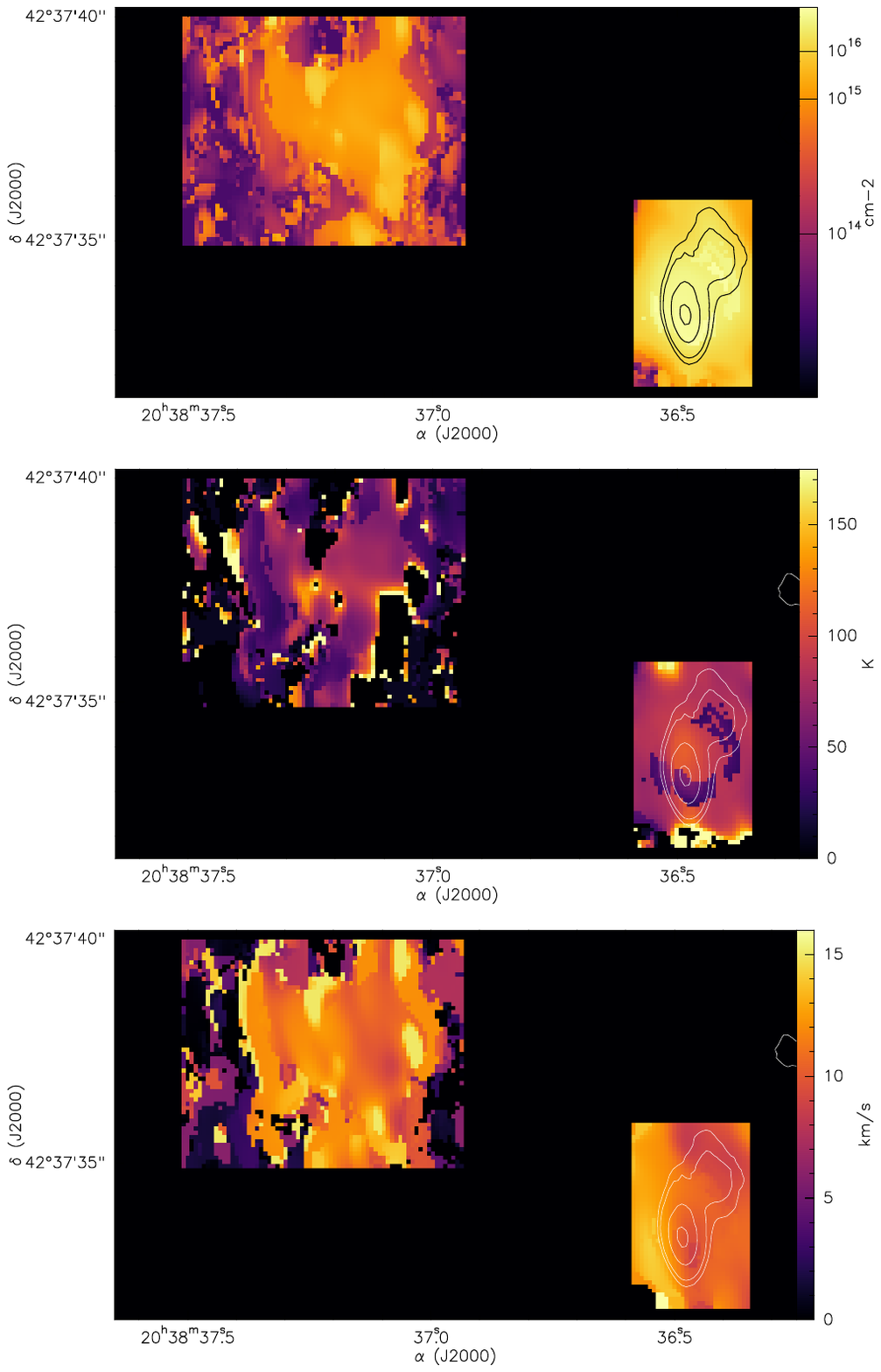}
\figsetgrpnote{Parameter maps for \ce{SO2} across MM1 and MM3 (from top to bottom: column density, kinetic temperature, and v$_{lsr}$). Black and white contours correspond to the continuum levels at 2, 3, 7, and 12 times $\sigma$ ($\sigma$ = 3.079 mJy beam$^{-1}$). The colors of the contours are arbitrary and differ between maps purely for aesthetic purposes. In the MM3 region, most pixels have uncertainties greater than the derived parameter values. The analysis of parameters across this region using this mapping method should therefore be seen as an initial estimate rather than a quantitative result.}
\figsetgrpend

\figsetend

\begin{figure}
\figurenum{15}
\plotone{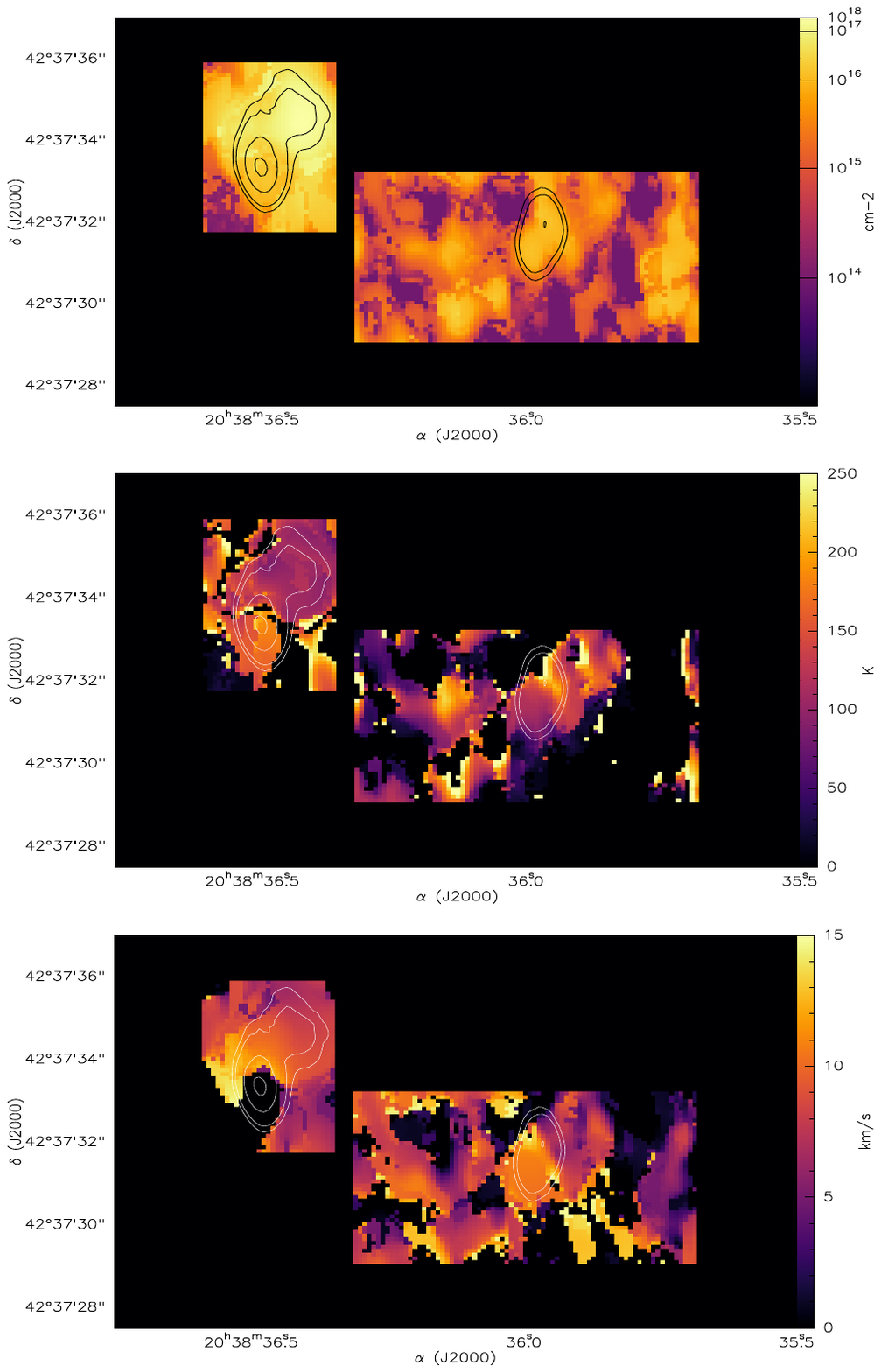}
\caption{\label{fig:parammaps}Parameter maps for \ce{CH3OCH3} across MM1 and MM2 (from top to bottom: column density, kinetic temperature, and v$_{lsr}$). Black and white contours correspond to the continuum levels at 2, 3, 7, and 12 times $\sigma$ ($\sigma$ = 3.079 mJy beam$^{-1}$). The colors of the contours are arbitrary and differ between maps purely for aesthetic purposes. In the MM2 region, most pixels have uncertainties greater than the derived parameter values. The analysis of parameters across this region using this mapping method should therefore be seen as an initial estimate rather than a quantitative result.}
\end{figure}

\section{Uncertainty Maps} \label{sec:appendix}

\begin{figure}
\figurenum{16}
\plotone{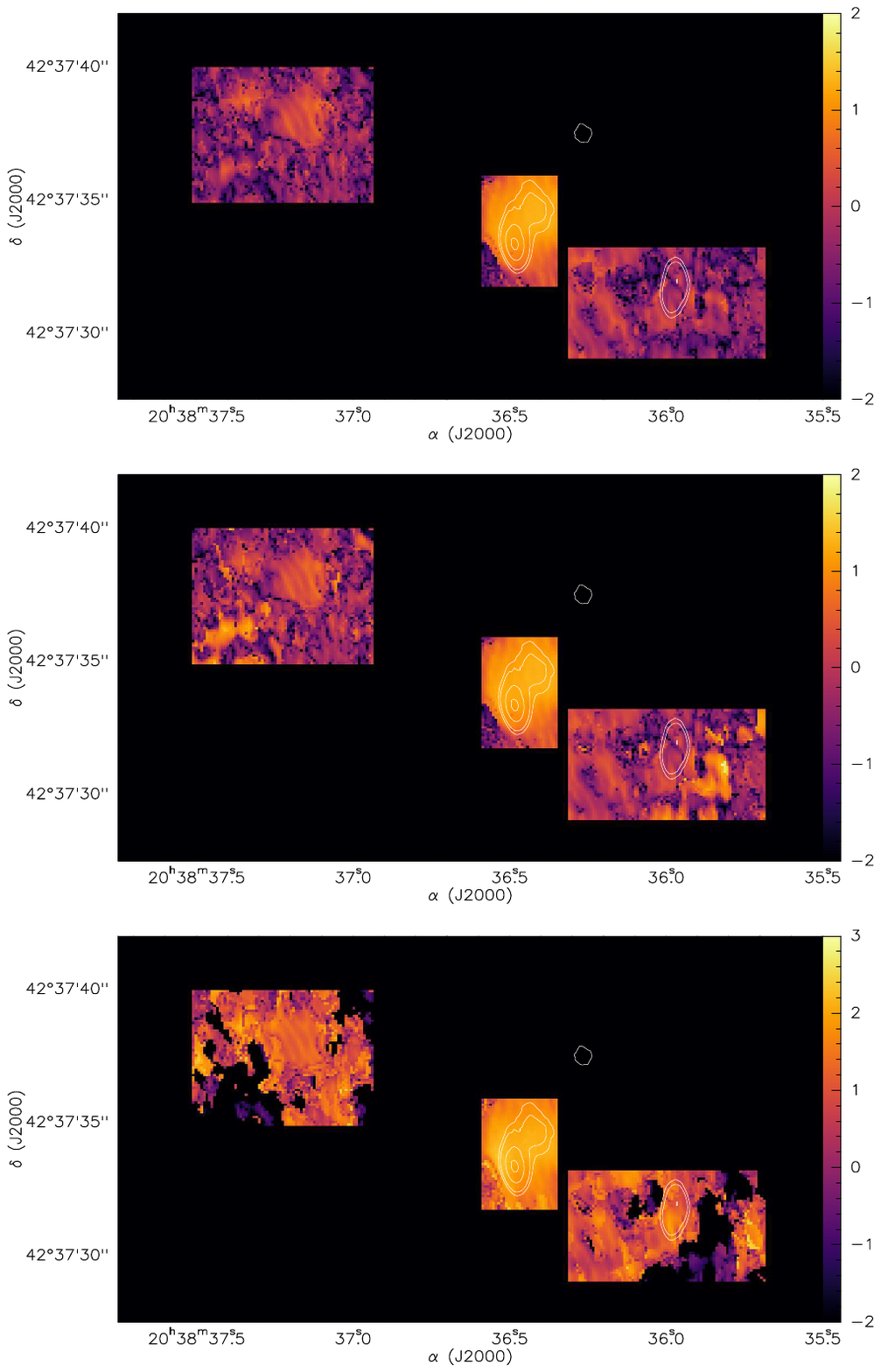}
\caption{\label{fig:uncertainty}Maps describing the uncertainty of parameter values for CH$_3$CN. See the \ref{sec:appendix} for an in-depth description of the map.}
\end{figure}

This section provides uncertainty values for the physical parameters derived using GOBASIC. For each individual pixel and parameter in Figure \ref{fig:CH3CN} and Figures \ref{fig:parammaps}, the derived values were divided by the corresponding uncertainty. Figures \ref{fig:uncertainty} plot these fractions spatially on a logarithmic scale with the more negative values in the plot corresponding to higher fractional uncertainty. The complete figure set (6 images) is available in the online journal.

\figsetstart
\figsetnum{16}
\figsettitle{Uncertainty Maps}

\figsetgrpstart
\figsetgrpnum{16.1}
\figsetgrptitle{\ce{CH3CN} Uncertainty}
\figsetplot{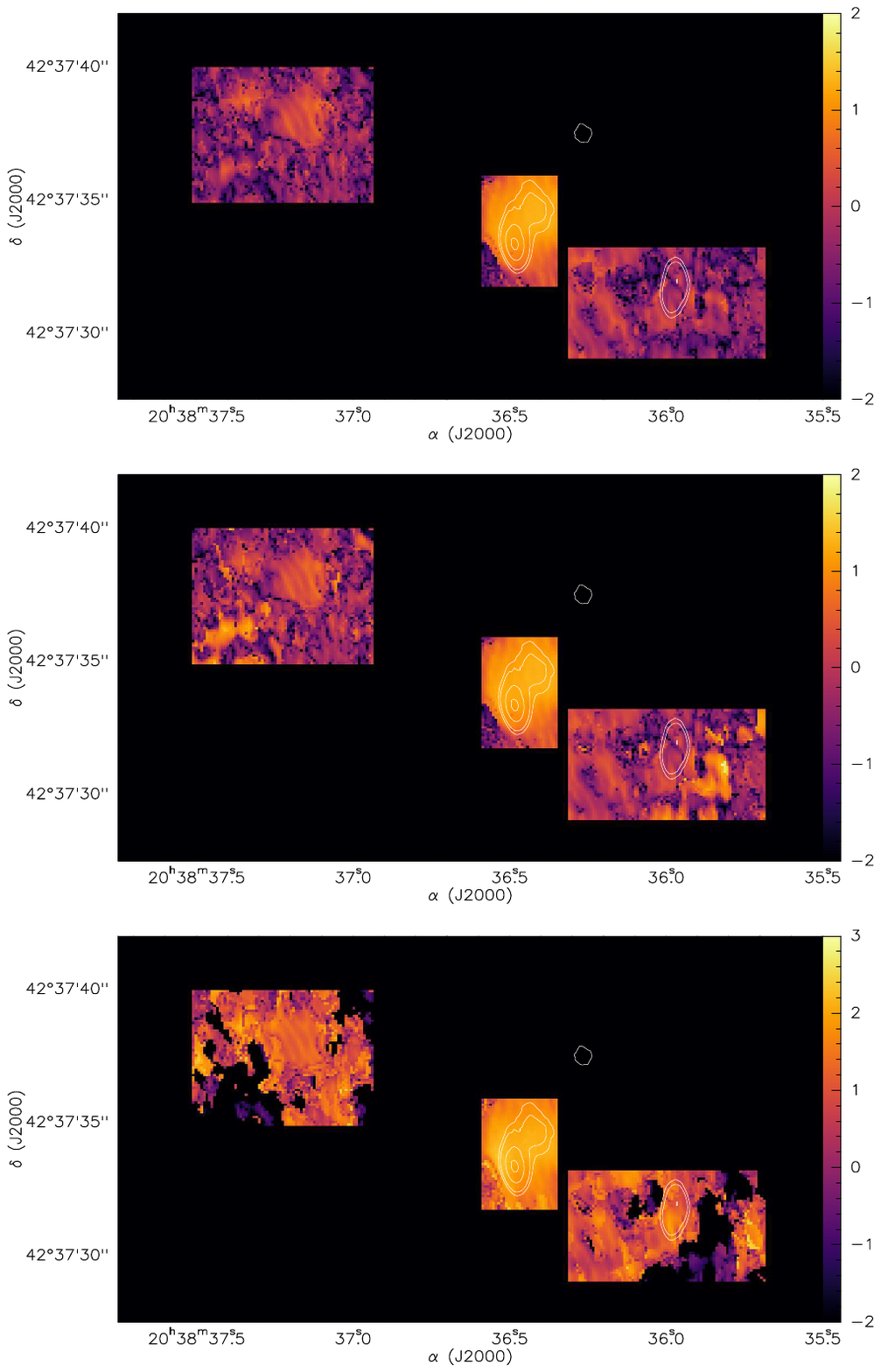}
\figsetgrpnote{Maps describing the uncertainty of parameter values for \ce{CH3CN}. See the \ref{sec:appendix} for an in-depth description of the map.}
\figsetgrpend

\figsetgrpstart
\figsetgrpnum{16.2}
\figsetgrptitle{\ce{CH3OCH3} Uncertainty}
\figsetplot{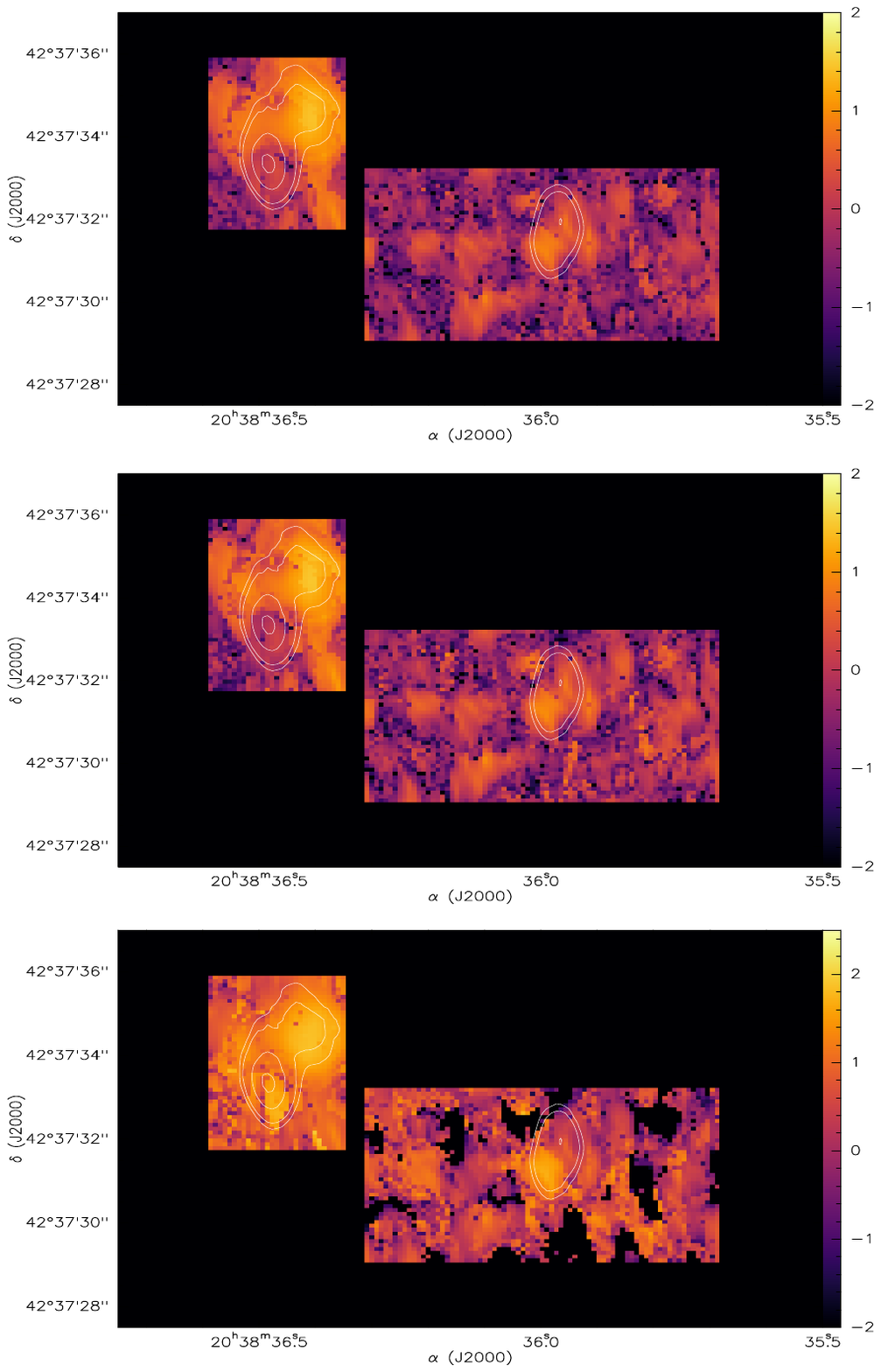}
\figsetgrpnote{Maps describing the uncertainty of parameter values for \ce{CH3OCH3}. See the \ref{sec:appendix} for an in-depth description of the map.}
\figsetgrpend

\figsetgrpstart
\figsetgrpnum{16.3}
\figsetgrptitle{\ce{CH3OH} Uncertainty}
\figsetplot{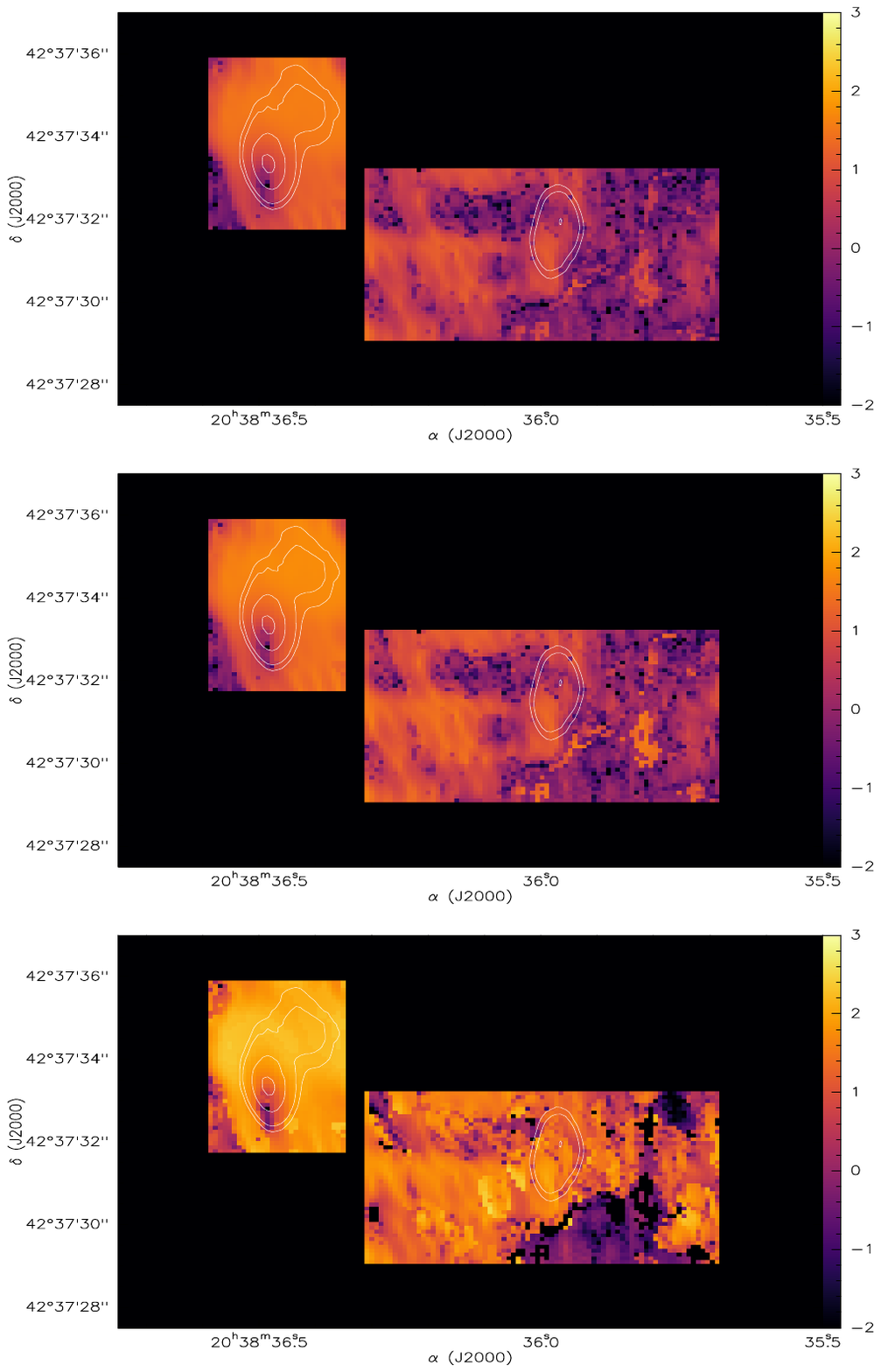}
\figsetgrpnote{Maps describing the uncertainty of parameter values for \ce{CH3OH}. See the \ref{sec:appendix} for an in-depth description of the map.}
\figsetgrpend

\figsetgrpstart
\figsetgrpnum{16.4}
\figsetgrptitle{\ce{HCOOCH3} Uncertainty}
\figsetplot{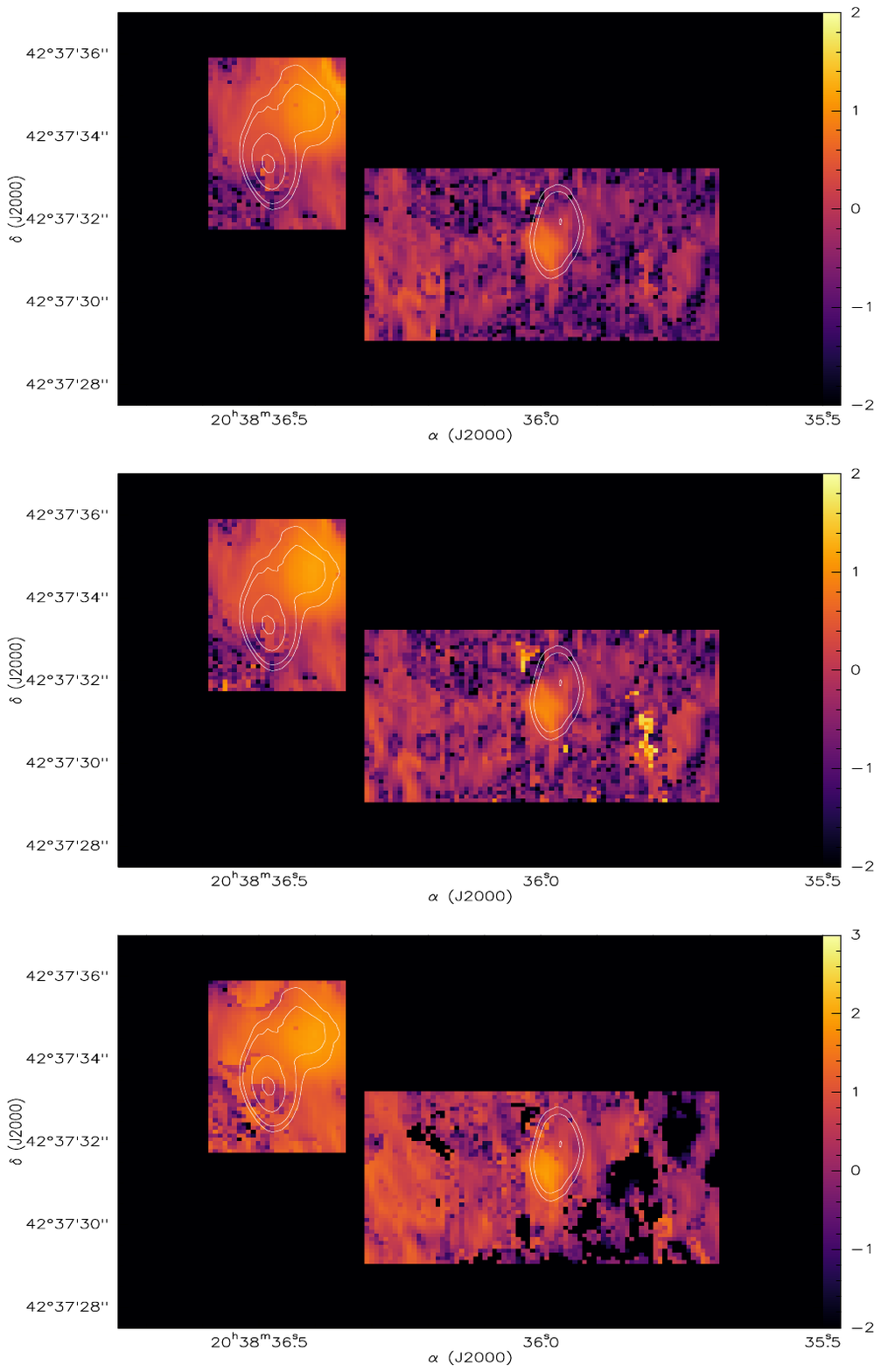}
\figsetgrpnote{Maps describing the uncertainty of parameter values for \ce{HCOOCH3}. See the \ref{sec:appendix} for an in-depth description of the map.}
\figsetgrpend

\figsetgrpstart
\figsetgrpnum{16.5}
\figsetgrptitle{\ce{HNCO} Uncertainty}
\figsetplot{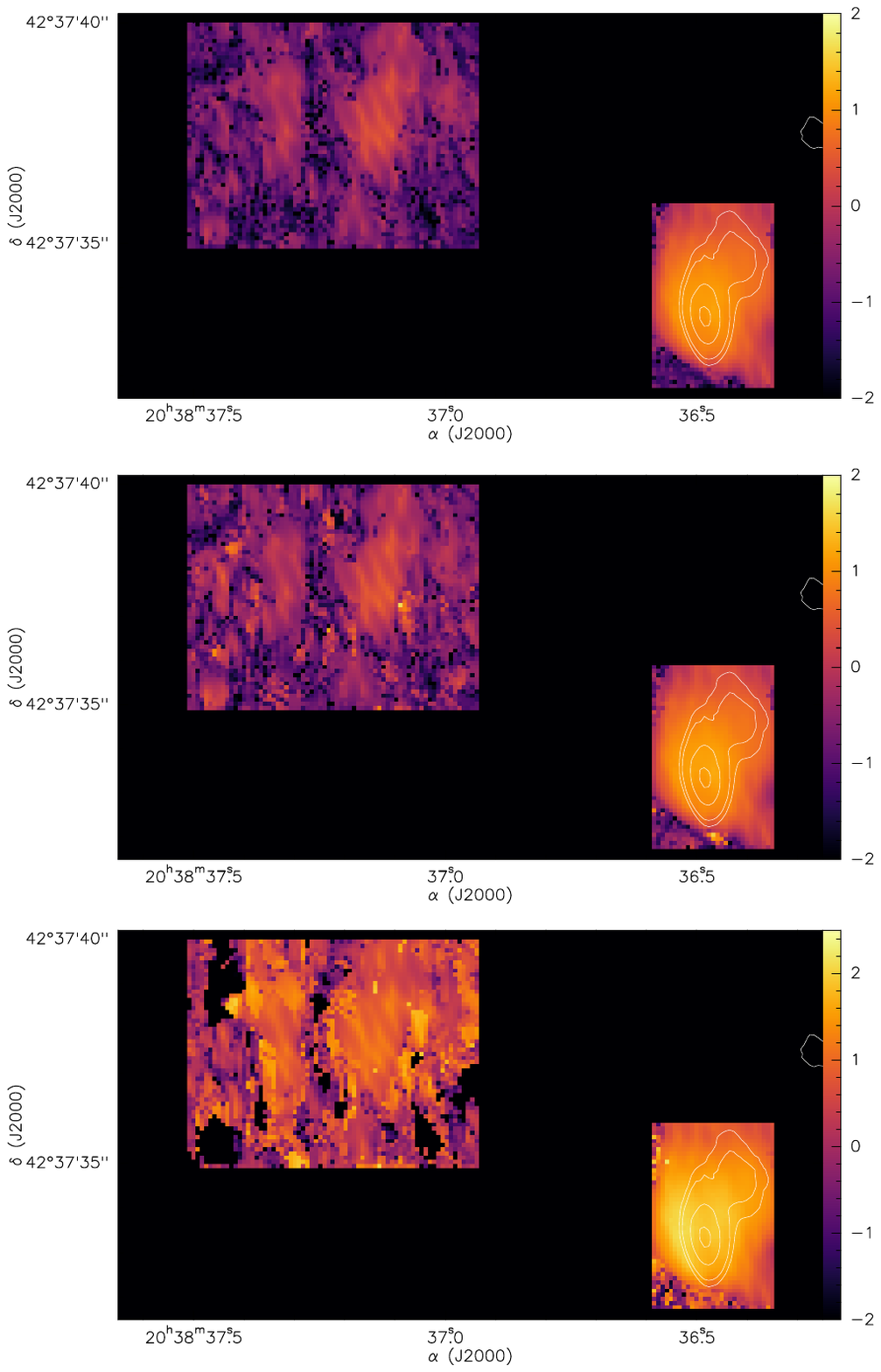}
\figsetgrpnote{Maps describing the uncertainty of parameter values for \ce{HNCO}. See the \ref{sec:appendix} for an in-depth description of the map.}
\figsetgrpend

\figsetgrpstart
\figsetgrpnum{16.6}
\figsetgrptitle{\ce{SO2} Uncertainty}
\figsetplot{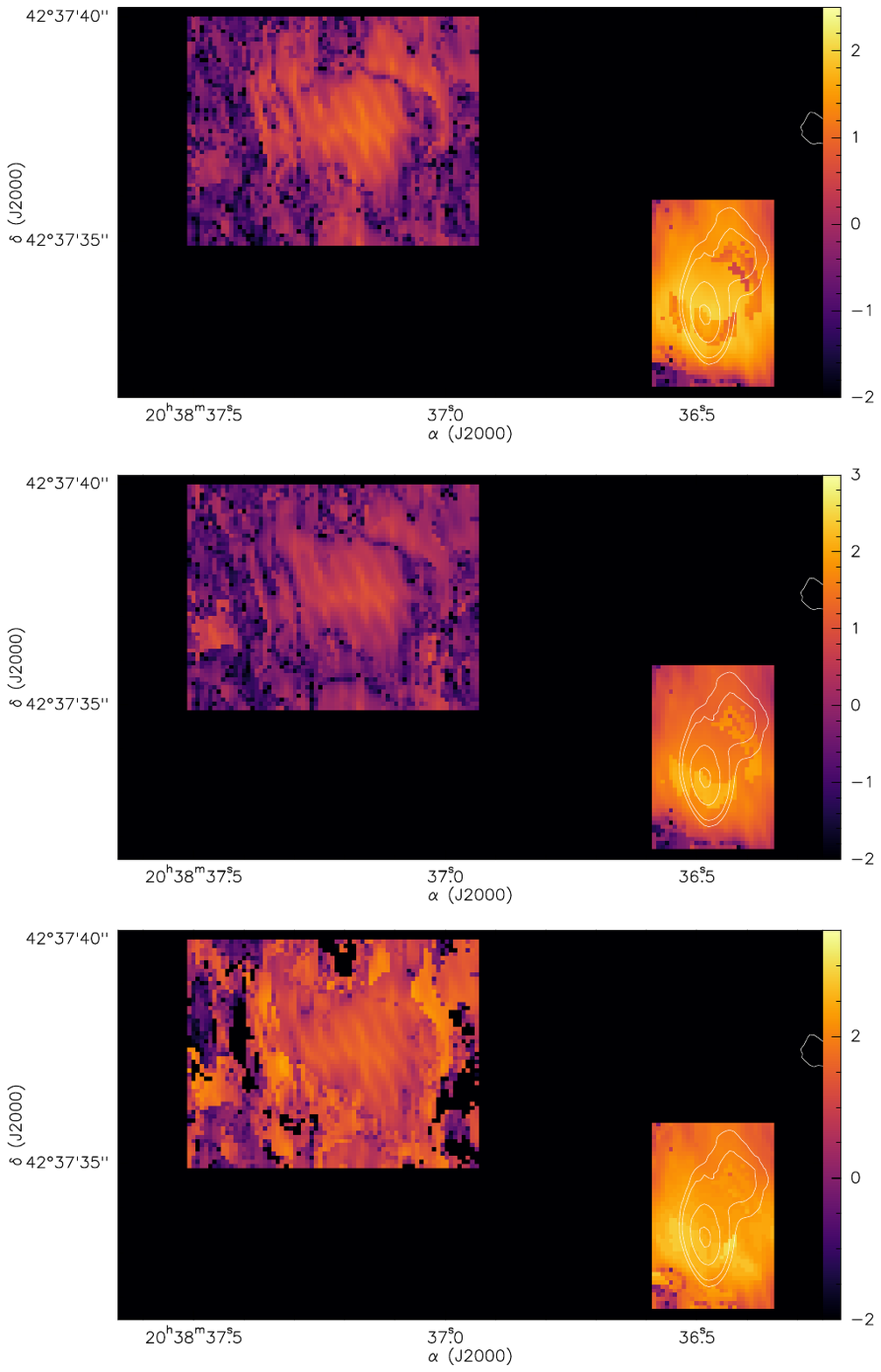}
\figsetgrpnote{Maps describing the uncertainty of parameter values for \ce{SO2}. See the \ref{sec:appendix} for an in-depth description of the map.}
\figsetgrpend

\figsetend

\end{appendix}

\bibliography{MyLibrary2,bib-doc}{}
\bibliographystyle{aasjournal}

\end{document}